\documentclass[twocolumn]{aa}  

\usepackage{graphicx}
\usepackage{txfonts}
\usepackage{color}
\usepackage{xcolor}
\usepackage{multicol}
\usepackage{amsmath}
\usepackage{makecell}
\usepackage{subcaption}
\usepackage{pdflscape}
\usepackage{longtable}
\usepackage{booktabs}
\usepackage{caption}
\usepackage[skip=2pt plus1pt, indent=15pt]{parskip}
\usepackage{threeparttablex}
\usepackage{tikz,caption}
\usepackage{hyperref}
\usetikzlibrary{calc}
\DeclareCaptionFormat{overlay}{\gdef\capoverlay{#1#2#3\par}}
\DeclareCaptionLabelSeparator{frcolon}{ $\,:$ }
\DeclareCaptionStyle{overlay}{format=overlay}
\usepackage{csvsimple}

\hypersetup{
    colorlinks=true,
    linkcolor=black,
    citecolor=blue,     
    urlcolor=magenta,
    linkbordercolor={red}
}

\begin{document}

   \title{Emergence of high-mass stars in complex fiber networks (EMERGE)}

   \subtitle{III. Fiber networks in Orion}

   \titlerunning{EMERGE III. Fiber networks in Orion}

   \author{A. Socci\inst{1}
          \and
          A. Hacar\inst{1}
          \and
          F. Bonanomi\inst{1}
          \and
          M. Tafalla\inst{2}
          \and
          S. Suri\inst{1}
          }

   \institute{1 - Institute for Astronomy (IfA), University of Vienna,
              T\"urkenschanzstrasse 17, A-1180 Vienna\\
              \email{andrea.socci@univie.ac.at} \\
              2 - Observatorio Astronómico Nacional (IGN), Alfonso XII 3, E-28014 Madrid, Spain
             }

   \date{Received; -- accepted; --}

 
  \abstract
   {The {\it Herschel} far infrared observations demonstrated the complex organisation of the interstellar medium in networks of parsec-scale filaments over the past decade. At the same time, networks of fibers have been recognised describing the gas structures in star-forming regions at sub-parsec scales.}
   {We aim to investigate the dense gas organisation prior to the formation of stars in a selected sample of regions within Orion.}
   {We surveyed seven prototypical star-forming regions in Orion as part of the EMERGE Early ALMA Survey. Our sample includes low- (OMC-4~South, NGC~2023), intermediate- (OMC-2, OMC-3, LDN~1641N), and high-mass (OMC-1, Flame Nebula) star-forming regions all surveyed at high spatial resolution (4.5\arcsec or $\sim$~2000~au) in N$_2$H$^+$ (1$-$0) using a dedicated series of ALMA+IRAM-30m observations. Using this homogeneous sample, we systematically investigated the spatial distribution, the density, and thermal structure of the star-forming gas, its column density variations, and its internal motions in a wide range of environments.
   }
   {From the analysis of the gas kinematics, we identified and characterised a total of 152 velocity-coherent fibers in our survey. The statistical significance of our sample, the largest of its kind so far, highlights these small-scale filamentary sub-structures as the preferred organisational unit for the dense gas in low-, intermediate- and high-mass star-forming regions alike. Despite the different complexity of these sub-parsec networks, in terms of surface density of their constituent fibers, the masses and lengths of these objects show similar distributions and consistent median values, as well as (trans-)sonic motions, in all of our targets.
   The comparison between the fiber line masses and virial line masses suggests the majority of these objects to be sub-virial. Those fibers closer to the virial condition, however, also have more protostars associated to them.
   Finally, the surface density of fibers is linearly correlated with the total dense gas mass throughout roughly one order of magnitude in both parameters.
   }
   {While most fibers show comparable mass, length, and internal motions in our survey, massive fibers, close to the virial condition, prove intimately connected to star formation. The majority of the protostars in our target regions are in fact associated to these objects. The additional correlation between the surface density of fibers and the dense gas mass in our survey demonstrates how fibers can explain the current star formation properties of their host region. These findings suggest a common mechanism for star formation from low- to high-mass star-forming regions mediated through the formation and evolution of fiber networks.}


   \maketitle
%
\renewcommand{\ttdefault}{pcr}
\captionsetup{labelfont=bf}

\section{Introduction}\label{sec:introduction}

Over the last century, a large variety of observations identified the filamentary structure of the interstellar medium (ISM) \citep{barnard07,lynds62,schneider79}. The proximity of the elongated structures to the young stellar objects in these regions suggested a connection between the two and to the star-formation process \citep[e.g.][]{hartmann02}. A large variety of parsec-scale filaments of gas and dust were observed in several star-forming regions since then \citep[e.g. integral shaped filament, ISF, in Orion;][]{bally87}. The advent of {\it Herschel} observations showed the widespread presence of filaments in the ISM and their role as drivers of star formation inside clouds \citep{andre14}.

A plethora of filamentary structures are nowadays identified in the ISM at all scales \citep{hacar22}. The far-infrared (FIR) dust emission observations from \textit{Herschel} probed filament networks at parsec scales both in nearby clouds \citep[e.g.][]{arzoumanian19} and Galactic Plane surveys \citep[e.g.][]{molinari10}. Large-scale surveys in dust extinction and emission identified giant filaments with sizes up to $\sim100$~pc \citep[e.g.][]{jackson10,goodman14}. Inside many of these parsec-size filaments, molecular line observations revealed the existence of velocity-coherent, sub-filaments \citep{hacar13}, usually referred to as fibers \citep{andre14}. These fibers have been identified both in single-dish \citep[e.g.][]{tafalla15,hacar17} and in interferometric studies \citep[e.g.][]{lee14,hacar18,dhab19} over the years. 
While the distinction between giant filaments, parsec-size filaments, and sub-parsec fibers is primarily driven by observations, all these filament families are recognised as part of the hierarchical structure of the ISM down to the $\sim1000$~au scale regime \citep{hacar22}.

Recognised as part of the hierarchy in the filamentary ISM, fibers play a pivotal role in the star formation process. They are in fact suggested to form when the turbulence characterising the ISM at large scales dissipates. Fibers would then set the initial conditions for core formation, which, in turn, would inherit their kinematic properties \citep{hacar11}. Despite the close connection to cores and, therefore, star formation, the search for velocity-coherent fibers is limited by our ability to resolve these structures, both in space and velocity. The need for observations at high spectral and spatial resolution has confined the study of fibers to single regions, thus limiting the overall statistics \citep[][and references therein]{hacar22}.

We aim to characterise the physical properties of velocity-coherent structures, and the potential variation of these properties with the environment, in a statistically significant sample of seven low- to high-mass regions in Orion. All our targets were homogeneously surveyed at high spatial resolution of $\sim2000$~au with the Atacama Large Millimetre Array (ALMA). This pilot study is part of the Emergence of high-mass stars in complex fiber networks (EMERGE) project \citep[see][hereafter Paper I]{hacar24}, which explores the origin of low- and high-mass stars in complex associations of filaments using ALMA observations. This Paper III in the series is organised as follows: we first present our sample (Sect.~\ref{sec:observations}) and the dense gas properties traced by our N$_2$H$^+$ observations (Sect.~\ref{sec:densegas}). We then describe our analysis, which focuses on the identification and characterisation of velocity-coherent structures in the regions included in the survey (Sect.~\ref{sec:analysis}). We explore the physical properties of these structures, recognising them as fibers, and we further assess their virial condition and their connection to the current star-formation in the host region (Sect.~\ref{sec:fibersurvey}). We finally conclude that fiber networks are the preferred organisation of the dense gas prior to the formation of stars in Orion. The complexity of these networks is mostly determined by the total dense gas mass available in the region, independently of its star-formation history or classification (e.g. low-, intermediate-, or high-mass). We summarise our conclusions and the main findings of the paper in Sect.~\ref{sec:conclusions}.

\begin{figure*}[tbp]
  \centering
  \captionsetup{format=overlay, labelsep=frcolon}
  \caption{}
  \begin{tikzpicture}
    \node[anchor=south west,inner sep=0] (image) at (0,0)
    {\includegraphics[width=0.98\linewidth]{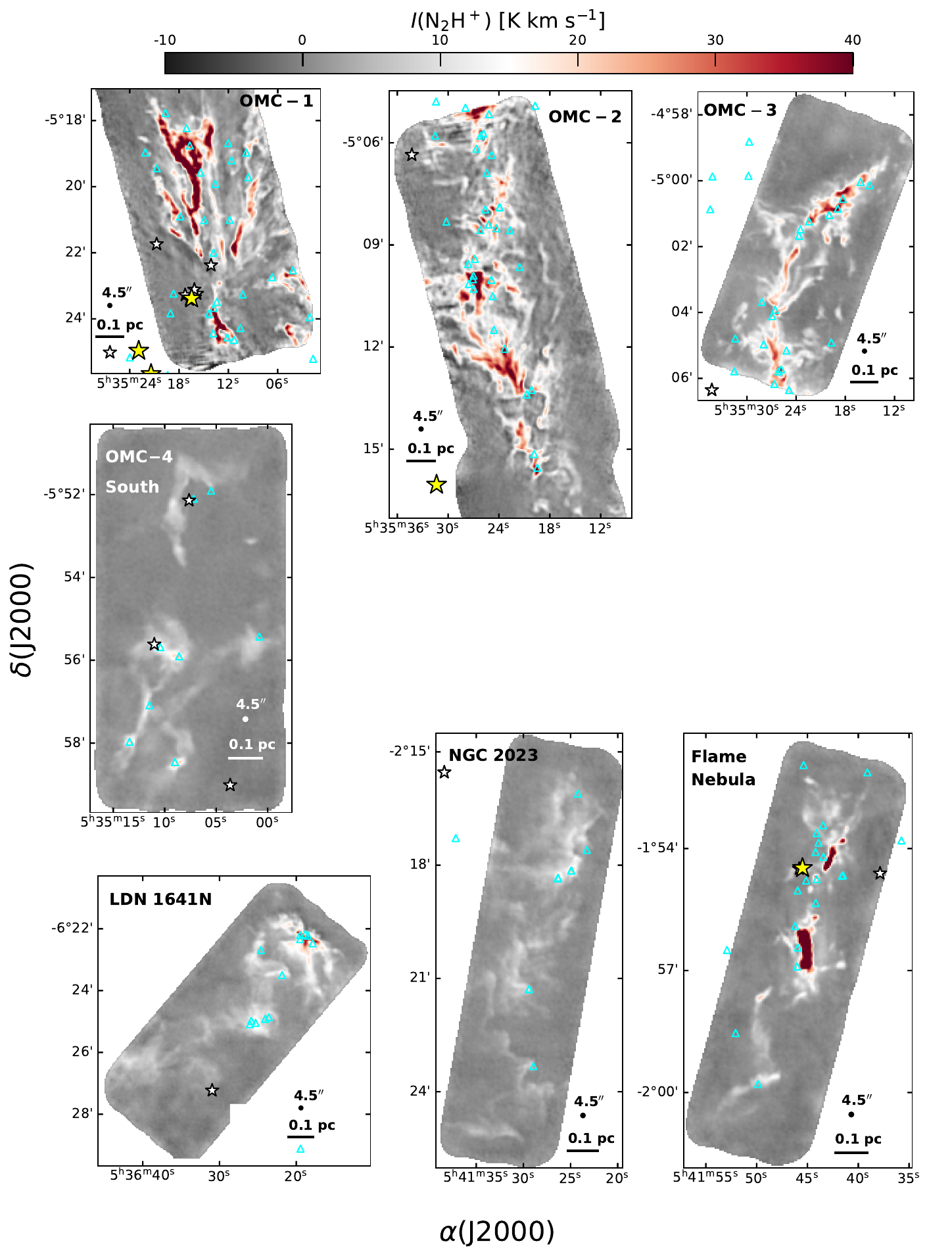}};
    \begin{scope}[x={(image.south east)},y={(image.north west)}]
      \draw let \p1 = (0.67,0.5)
      in node[text width=\x1,align=center,color=black] at
      (0.67, 0.5) {\textbf{Fig.~1:} N$_2$H$^+$ (1$-$0) integrated intensity maps of the seven star-forming regions in the EMERGE Early ALMA Survey at 4.5$''$ resolution with ALMA+IRAM-30m. In all panels, cyan triangles are the protostellar objects from \protect\citet{megeath12,stutz13,furlan16}, while the yellow and white stars are respectively the O-B stars collected from Simbad \protect\citep{wenger00}.};
    \end{scope}
  \end{tikzpicture}
  \label{fig:In2h+}
\end{figure*}

\begin{figure*}[tbp]
  \centering
  \captionsetup{format=overlay, labelsep=frcolon}
  \caption{}
  \begin{tikzpicture}
    \node[anchor=south west,inner sep=0] (image) at (0,0)
    {\includegraphics[width=0.98\linewidth]{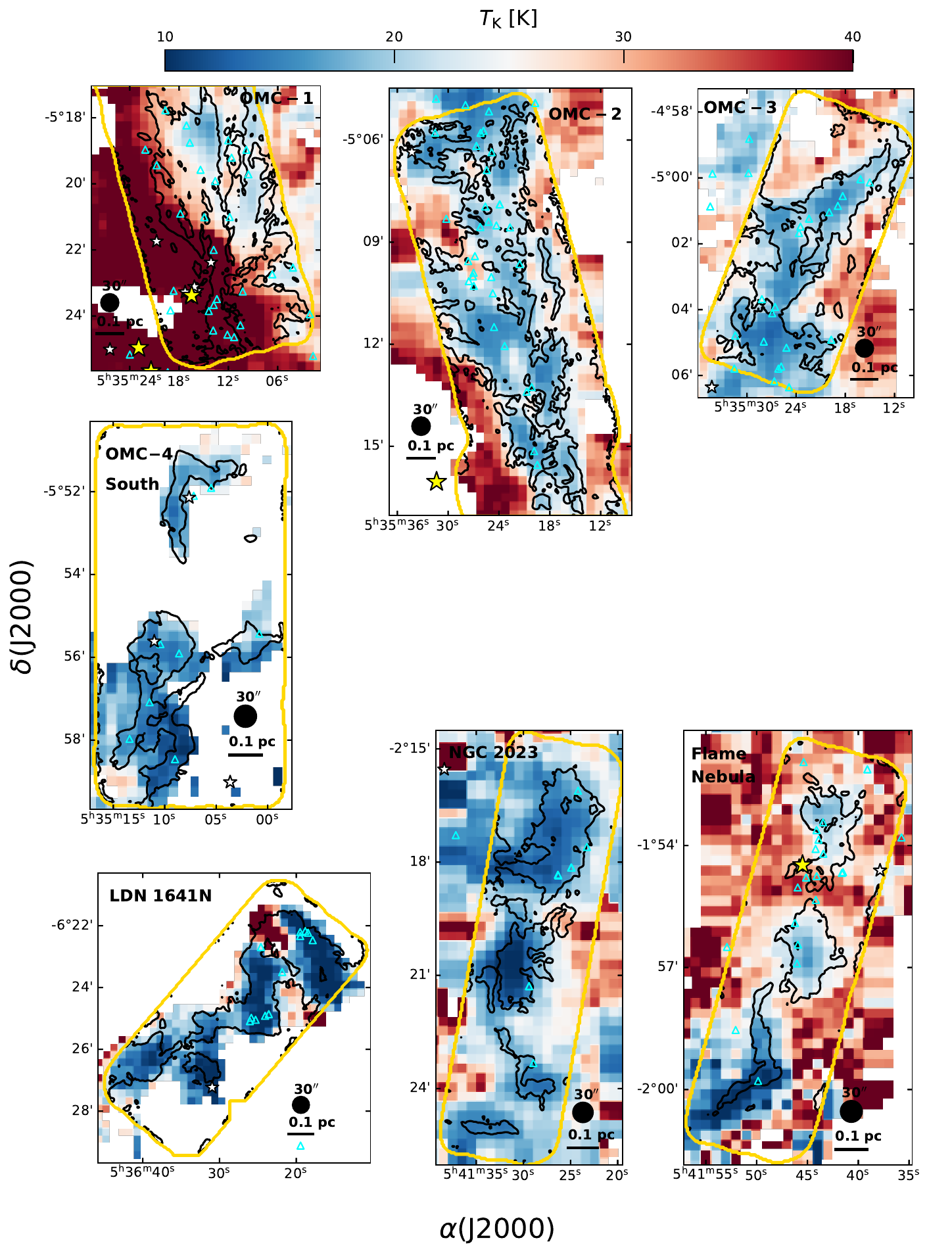}};
    \begin{scope}[x={(image.south east)},y={(image.north west)}]
      \draw let \p1 = (0.67,0.5)
      in node[text width=\x1,align=center,color=black] at
      (0.67, 0.5) {\textbf{Fig.~2:} Temperature maps derived from the HCN-to-HNC ratio at 30$''$ with IRAM-30m (see Paper I). The fields with $I(\mathrm{HNC}) < 2$~K~km~s$^{-1}$ have been masked. The black contours show $I(\mathrm{N_2H^+})\geq 2.5~\mathrm{K~km~s}^{-1}$, three times the noise estimated from the emission-free sub-regions in the maps.
    Same as Fig.~\ref{fig:In2h+}, cyan triangles are the protostellar objects while the yellow and white stars are the O-B stars, respectively.}; 
    \end{scope}
  \end{tikzpicture}
  \label{fig:Tk}
\end{figure*}

\begin{figure*}[tbp]
  \centering
  \captionsetup{format=overlay, labelsep=frcolon}
  \caption{}
  \begin{tikzpicture}
    \node[anchor=south west,inner sep=0] (image) at (0,0)
    {\includegraphics[width=0.98\linewidth]{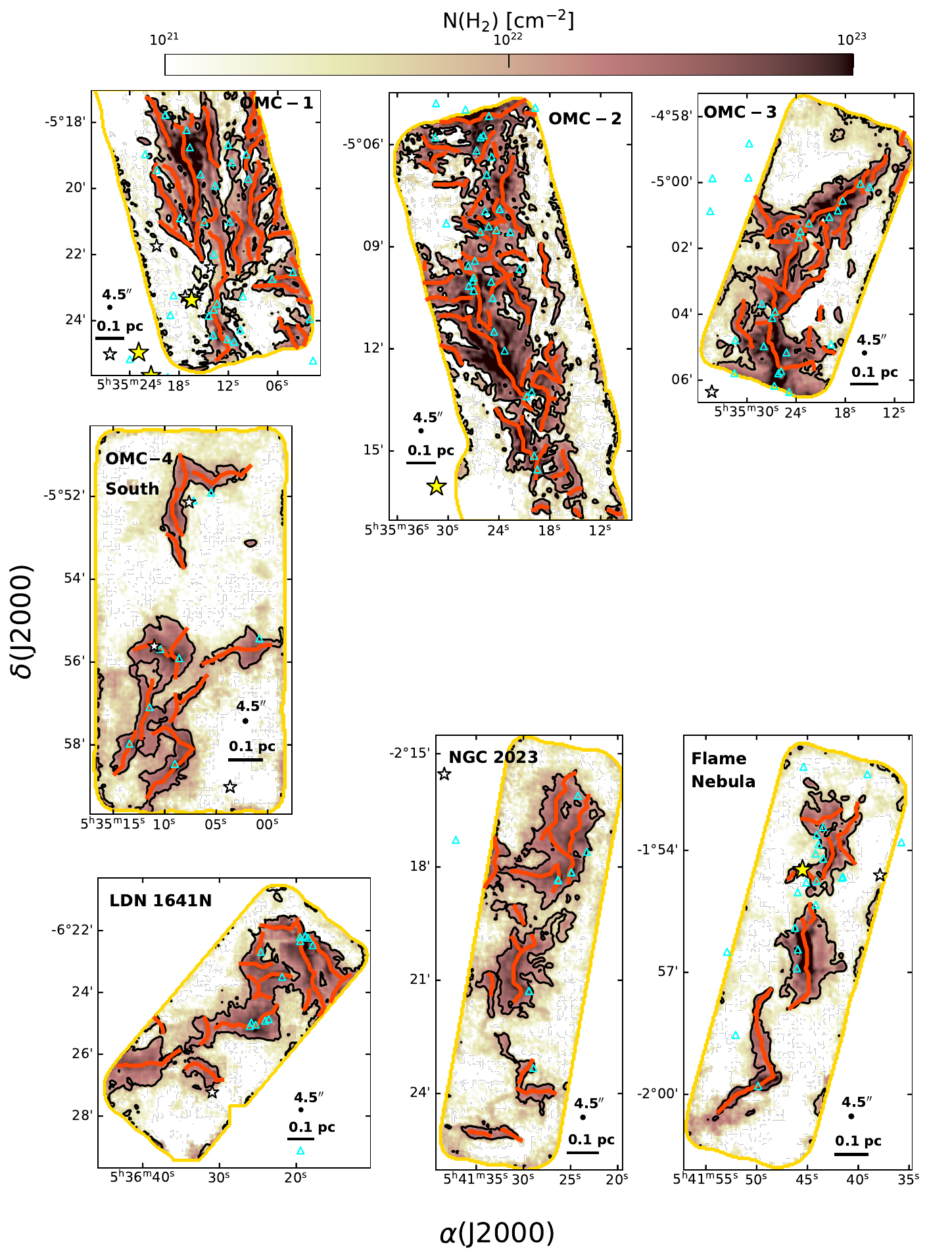}};
    \begin{scope}[x={(image.south east)},y={(image.north west)}]
      \draw let \p1 = (0.67,0.5)
      in node[text width=\x1,align=center,color=black] at
      (0.67, 0.5) {\textbf{Fig.~3:} Total column density maps of the regions derived from the temperature corrected N$_2$H$^+$ (1$-$0) integrated intensity (see Eq.~\ref{eq:caln2hp}). The black contour corresponds again to an intensity of N$_2$H$^+$ three times the noise. The red lines represent instead the axes of the fibers identified in our kinematic analysis (see Sect.~\ref{sec:analysis}). Same as Fig.~\ref{fig:In2h+}, cyan triangles are the protostellar objects while the yellow and white stars are respectively the O-B stars.};
    \end{scope}
  \end{tikzpicture}
  \label{fig:NH2}
\end{figure*}

\section{EMERGE Early ALMA Survey}\label{sec:observations}

\begin{table*}[tbp]
    \caption{General properties of the EMERGE Early ALMA Survey.}
    \centering
    \begin{tabular*}{\textwidth}{c@{\hspace{2.5\tabcolsep}}c@{\hspace{2\tabcolsep}}c@{\hspace{2\tabcolsep}}c@{\hspace{3\tabcolsep}}c@{\hspace{3\tabcolsep}}c@{\hspace{2\tabcolsep}}c@{\hspace{2\tabcolsep}}c@{\hspace{2\tabcolsep}}c@{\hspace{2\tabcolsep}}c@{\hspace{2.5\tabcolsep}}}
    \hline
    Source & Cloud & $D$ [pc] & SF-regime & O-star & Evolutionary stage & P\tablefootmark{*} & P+D\tablefootmark{*} & P/D\tablefootmark{*} & ALMA Proj.ID \\
    \hline \\ [-2ex]
    OMC-1 & Orion A & 400 & high- & Yes & Evolved & 36 & 423 & 0.09 & 2015.1.00669.S \\ [0.7ex]
    OMC-2 & Orion A & 400 & intermediate- & No & Young & 33 & 142 & 0.30 & 2015.1.00669.S \\ [0.7ex]
    OMC-3 & Orion A & 400 & intermediate- & No & Young & 26 & 102 & 0.34 & 2019.1.00641.S \\ [0.7ex]
    OMC-4 South & Orion A & 400 & low- & ? & Evolved & 11 & 59 & 0.23 & 2019.1.00641.S \\ [0.7ex]
    LDN~1641N & Orion A & 400 & intermediate- & No & Young & 13 & 51 & 0.34 & 2019.1.00641.S \\ [0.5ex] \hline \\ [-2ex]
    NGC~2023 & Orion B & 423 & low- & No & Young & 6 & 8 & 0.75 & 2019.1.00641.S \\ [0.7ex]
    Flame Nebula & Orion B & 423 & high- & Yes & Evolved & 21 & 141 & 0.18 & 2019.1.00641.S \\ [0.5ex]
    \hline
    \end{tabular*}
    \tablefoot{
    \tablefoottext{*}{The number of protostars (P) and disks (D) here listed are merely used as reference in this paper, as they refer to larger footprints compared to our ALMA fields (see Paper I for a full discussion).}
    }
    \label{tab:sample_prop}
\end{table*}

The EMERGE project aims to study the formation mechanism of high- and low-mass stars in densely populated fiber associations (i.e. fiber networks; see Paper I). While filament networks are reported on parsec-scale \citep[e.g.][]{andre14}, fiber networks are identified at sub-parsec scales describing the organisation of dense, star-forming gas prior to the formation of stars \citep{hacar18}. The EMERGE project will make use of both new and archival ALMA interferometric observations at high spatial resolution to resolve these fiber networks across the Milky Way. 

In this first pilot study, the EMERGE Early ALMA Survey explores seven star-forming regions in Orion, namely the Orion Molecular Clouds (OMCs) OMC-1, OMC-2, OMC-3, OMC-4 South, LDN~1641N in Orion A and NGC~2023, Flame Nebula (or NGC~2024) in Orion B. The targets of this sample, comprising high- (OMC-1, Flame Nebula), intermediate- (OMC-2, OMC-3, LDN~1641N) and low-mass (NGC~2023, OMC-4 South) star-forming regions, were chosen to cover different mass reservoirs, evolutionary stages, and stellar populations (see properties in Table \ref{tab:sample_prop}; see also Paper I for a discussion). 
The entire EMERGE Early ALMA sample was homogeneously surveyed at a resolution of 4.5$''$ \citep[or $\sim2000$ au at 414 pc;][]{menten07} using ALMA observations at 3mm (Band 3; Proj.IDs: 2015.1.00669.S, 2019.1.00641.S). 
All targets were observed using independent, large-scale ALMA mosaics, between 110 and 145 pointings each, typically covering an area of $\sim200\times600$~arcsec$^2$ (or $0.4\times1.2$~pc$^2$) in size.
Our ALMA data include independent observations of N$_2$H$^+$ (1$-$0) (93.17 GHz), HNC (1$-$0) (90.66 GHz), and HC$_3$N (10$-$9) (90.97 GHz), all observed at high spectral resolution ($\delta V\leq 0.15$~km~s$^{-1}$), with the addition of the 3mm-continuum. This suite of tracers was selected to probe different density regimes and physical processes within the sample (see also Paper I).
These interferometric ALMA observations were combined with additional low spatial resolution, IRAM-30m (single-dish) data, used as short-spacing information \citep[see][hereafter Paper II]{bonanomi24}. The final data products are high-sensitivity, high-resolution ALMA+IRAM-30m spectral cubes and continuum maps, whose fields are homogeneously reduced throughout our survey.
The ALMA+IRAM-30m data are complemented by ancillary catalogues of young stellar objects \citep[YSOs including Class 0/I prostostars and Class II disks;][]{megeath12,furlan16,stutz13}, and O-B stars \citep{wenger00}, and by {\it Herschel} FIR measurements (36$''$) of the total gas column density \citep[N(H$_2$);][]{lombardi14}, and single-dish maps (30$''$) of the gas kinetic temperature \citep[$T_\mathrm{K}$;][]{hacar20}.
We refer the reader to Paper I for a full description of both the new and ancillary data used here.

Among the different molecular species included in the EMERGE Early ALMA survey, we investigated here the N$_2$H$^+$ (1$-$0) emission as probe of the dense gas structure within our sample. N$_2$H$^+$ is a density selective tracer formed at densities $n(\mathrm{H_2})\gtrsim5\times10^4~\mathrm{cm}^{-3}$ when CO disappears from the gas phase due to freeze-out \citep[see][for an overview]{tafalla23}. Thanks to its favourable observational properties, several studies used the ground (J=1$-$0) transition of N$_2$H$^+$ as standard tracer of dense cores at sub-parsec scales in low-mass star-forming regions \citep[e.g.][]{caselli02}. The advent of high-sensitivity observations in recent years revealed the presence of extended N$_2$H$^+$ (1$-$0) emission tracing the filamentary structure of intermediate- and high-mass star-forming regions at parsec scales \citep{lee14,henshaw14,hacar18,chen19,barnes21}. 
In these studies, the N$_2$H$^+$ emission samples the high-density material in the regions, typically found at column densities $\gtrsim10^{22}$~cm$^{-2}$ \citep{bergin07}, and closely connected to the location of young protostars \citep[e.g.][]{lee14,hacar18,chen19}.
The study of the N$_2$H$^+$ (1$-$0) emission allowed for the systematic investigation of the dense gas physical properties (structure, kinematics, stability) prior to the formation of stars in low- to high-mass star-forming regions.

In the following sections, we explore the emission properties of N$_2$H$^+$ (1$-$0), homogeneously sampled by the corresponding ALMA+IRAM-30m mosaics towards the seven star-forming regions included in the EMERGE Early ALMA Survey. These seven mosaics are reduced using the Model-Assisted Cleaning plus Feather (MACF) method, recovering $\gtrsim90\%$ of the emission detected in our single-dish data throughout the survey (see Paper I for additional details). 
Each of these seven Nyquist-sampled ALMA fields is a spectral cube at spatial resolution of $\sim2000$~au, containing between $\sim18,000$ and 26,000 spectra observed at high-spectral resolution. 
Our analysis considers a total of $\sim$~170,000 spectra overall, becoming one of the largest ALMA surveys in N$_2$H$^+$ so far. Our cubes are originally reduced in flux units of Jansky~beam~$^{-1}$, then converted to Kelvin in main beam temperature units ($T_\mathrm{mb}$) using a standard conversion factor of 6.96~Jy~beam~$^{-1}$~K$^{-1}$ at a resolution of 4.5$''$ (see ALMA Technical Handbook\footnote{\url{https://almascience.eso.org/proposing/technical-handbook}}).

\section{Dense gas structure at 2000 au resolution}\label{sec:densegas}

The characterisation of the gas in the EMERGE Early ALMA Survey goes through the analysis of its kinematics, using the N$_2$H$^+$ spectra, but also through the determination of its physical properties, such as its total mass and the size of the substructures. In the following sections, we explore the morphology of the N$_2$H$^+$ emission (Sect.~\ref{subsec:n2h+}), the temperature regime covered by our observations (Sect.~\ref{subsec:tk}), and the combination of the two to derive high-resolution N(H$_2$) maps of our regions (Sect.~\ref{subsec:densegasprop}). We finally discuss how accurate is the EMERGE Early ALMA Survey in the description of the gas content compared to previous surveys in the same regions (Sect.~\ref{subsec:comparsurv}). 

\subsection{N$_2$H$^+$ emission}\label{subsec:n2h+}

We start our study from the global emission properties of N$_2$H$^+$ (1$-$0). Figure \ref{fig:In2h+} shows the integrated emission of N$_2$H$^+$ in each of the seven ALMA fields part of the survey. The integration range in velocity was adapted to the individual cloud velocities in each case. Superposed to these maps, we display the location of the individual protostars (cyan triangles), of the B- (white stars) and O-stars (yellow stars) stars identified in previous surveys within these regions. At a first glance, the sample presents a large variety of emission features, pairing bright peaks to regions with extended emission, all accurately recovered by the ALMA+IRAM-30m data combination. 

The coverage of our ALMA maps was chosen to follow the distribution of the N$_2$H$^+$ (1$-$0) emission and the high column density material ($\mathrm{N(H_2)}\gtrsim10^{22}$~cm$^{-2}$), both previously covered by low-resolution surveys (see Paper I for a description). At the low spatial resolutions ($30-36''$ or $\sim0.06$~pc) of these ancillary surveys, our targets appear as elongated parsec-size filaments in line and continuum emission alike.
When observed at high resolution with ALMA (4.5$''$ or $\sim0.01$~pc), these same regions show a more diverse morphology instead. 
Regions such as OMC-2 and OMC-3 break down into a plethora of high-contrast, sharp, and elongated emission features, roughly aligned with the large-scale structure of the cloud. Others, such as OMC-4 South and NGC~2023, appear more irregular and diffuse. Cases such as LDN~1641N and the Flame Nebula display a cometary shape. Finally, OMC-1 displays a convergent, hub-like \citep{myers09} distribution of structures. Some of these morphological differences were already documented in previous studies \citep[e.g.][]{wise98,johnstone06,mairs16,hacar18,stanke22} and denote the need for high spatial resolution observations when describing the dense gas structure of star-forming regions such as those in our survey.

In addition to these large-scale differences, the ALMA+IRAM-30m maps, displayed in Fig.~\ref{fig:In2h+}, show a large variation in the N$_2$H$^+$ (1$-$0) intensity. The brightest integrated emission peaks are found in the Flame Nebula (125~K~km~s$^{-1}$), OMC-1 (88~K~km~s$^{-1}$), OMC-2 (77~K~km~s$^{-1}$), and LDN~1641N (77~K~km~s$^{-1}$). These peaks usually correspond to the most active star-forming sites in these fields, as denoted by the presence of multiple protostars and highly embedded sources detected in the millimeter/FIR continuum (cyan triangles in our images). That is the case for the OMC-1 Ridge \citep{teixeira16}, OMC-1 South \citep{rivilla13,palau18}, OMC-2 FIR-4, OMC-3 MMS sources \citep[see][for a review]{peterson08} and the head of LDN~1641N \citep{megeath16}. The bright emission detected in these active star-forming sites contrasts with the much weaker emission seen in more diffuse clouds, such as OMC-4~South and NGC~2023 (both with intensity peaks of $\sim21$~K~km~s$^{-1}$). The relative variation seen in these peaks is usually accompanied by a proportional amount of diffuse and extended emission, seen down to 2.5~K~km~s$^{-1}$. In all cases, the emission is confidently detected above the noise level our of maps, estimated within $0.4-1.1$~K~km~s$^{-1}$ (depending on the field) from the analysis of different emission-free sub-regions in Fig.~\ref{fig:In2h+}. With a dynamic range between 25 (OMC-4) and 255 (Flame Nebula), our ALMA+IRAM-30m maps are able to capture two orders of magnitude in N$_2$H$^+$ intensity, describing both its compact and extended emission components in these clouds. 

While exploring a large range of physical conditions, our ALMA+IRAM-30m maps display surprising similarities when resolved at 2000~au resolution. In all of our targets the N$_2$H$^+$ emission is heavily structured forming a series of elongated filamentary features with sizes on the order of $\lesssim0.1-0.2$~pc.
These small-scale filaments also have widths well below 0.1~pc in most cases, and close to the resolution of our observations (see beam size and scale bar in our figures). Far from idealised cylinders, these small filaments have a high internal variability with multiple branches, frequent changes in intensity, and variable width across their axis. 
This fibrous substructure (see Sect.~\ref{subsec:densegasprop} for a further discussion) was previously reported in the ALMA observations of N$_2$H$^+$ (1$-$0) towards OMC-1 and OMC-2 alone \citep{hacar18}. The same filamentary substructure seen throughout our survey suggests these features as intrinsic of the dense gas organisation in our regions, independently of their star-formation regime (low- vs high-mass), evolutionary stage (young vs evolved), or stellar content (with or without O-stars).

While common to all of our targets, the arrangement of these filamentary substructures seen in N$_2$H$^+$ (1$-$0) integrated intensity may change across our sample. Regions with large (proto-)stellar populations (see triangles and stars in Fig.~\ref{fig:In2h+}), such as OMC-1, OMC-2, OMC-3, or the northern end of LDN~1641N, show highly concentrated structures organised in densely populated networks of filaments at sub-parsec scales. At the location of the brightest emission peaks in our maps (see above), multiple filament junctions and condensations with sizes of $\lesssim$~0.1~pc are recognised. These condensations correspond to previously reported sources, such as OMC-1 Ridge, OMC-1 South, OMC-2 FIR-4 and FIR-6, OMC-3 MMS8-9, MMS1-5, and LDN~1641N.
In contrast, regions with shallow (proto-)stellar populations, such as OMC-4~South and NGC~2023, or under the influence of strong feedback, such as the Flame Nebula, display a more diffuse and distributed substructure.

Before continuing our study, it is important to investigate which gas conditions are sampled by our N$_2$H$^+$ observations. From the analysis of single-dish (30$''$) observations, Paper I determined a strong correlation between the location of the bright N$_2$H$^+$ (1$-$0) emission and the position of the (Class 0/I, $\sim0.5$~Myr old) protostars reported in previous IR surveys (cyan triangles; P in Table~\ref{tab:sample_prop}). This correlation is however lost when comparing the N$_2$H$^+$ emission with the positions of the (Class II/III, $\sim2$~Myr old) disks reported in those same surveys (not shown; D in Table~\ref{tab:sample_prop}). The ratio P/D is a classical evolutionary tracer \citep{evans09}: young regions are expected to have P/D~$\gtrsim1$. Regions with continuous star-formation over time approach the steady-state value given by the typical lifetimes of the Class I/II objects (i.e. P/D~$\sim\frac{0.5~\mathrm{Myr}}{2.0~\mathrm{Myr}}\sim0.25$). More evolved regions harbour a larger number of Class II objects showing P/D~<~0.25 (see Paper I for a full discussion). From our ALMA+IRAM-30m maps, we determined most of the protostars in our fields to be located on top of (or close to) local N$_2$H$^+$ (1$-$0) intensity peaks, similarly to Paper I despite the difference in resolution (4.5$''$ vs 30$''$). This correlation is particularly strong in regions with a limited presence of O-B stars and P/D ratios $\gtrsim0.25$, such as OMC-2 or OMC-3. Exceptions to this general trend are instead found in the Flame Nebula and (part of) OMC-1, where the strong effects of UV-radiation \citep[as well as the incompleteness of the IR surveys in these bright regions, see][]{megeath16} might disrupt this correlation. This tight correspondence is expected since N$_2$H$^+$ is found to trace the fresh dense material with a typical lifetime of $\lesssim$~1~Myr, and currently forming stars with a roughly constant star-formation rate \citep{Priestley2023}. The total mass of the dense gas probed by N$_2$H$^+$ is therefore the major contributing factor to the star-formation activity in the regions (see Paper I for a further discussion). 

\subsection{Gas kinetic temperature}\label{subsec:tk}

Additional insights are gained by exploring the temperature regimes in which the N$_2$H$^+$ (1$-$0) emission is present. Figure~\ref{fig:Tk} shows the gas kinetic temperature ($T_\mathrm{K}$) obtained using IRAM-30m observations at 30$''$ resolution of the HCN (1$-$0) and HNC (1$-$0) lines (see Paper I for additional details). Their integrated intensity ratio, sensitive to temperature variations due to the formation and destruction pathways of the two species \citep[e.g.][]{herbst00}, was calibrated against independent $T_\mathrm{K}$ measurements obtained with NH$_3$ by \citet{hacar20}. By applying the empirical correlation determined by Hacar et al., we derived large scale temperature maps for our targets, which are most accurate within the range $T_\mathrm{K}\sim15-40$~K.

We found strong similarities between the morphology of the N$_2$H$^+$ (1$-$0) emission and the coldest regions in our targets overall. 
The bulk of the N$_2$H$^+$ (1$-$0) emission in our maps (see first contour superposed to our $T_\mathrm{K}$ maps) is usually found at temperatures between $T_\mathrm{K}\sim$~10~K (e.g. OMC-4 South) and $T_\mathrm{K}\lesssim$~25~K (e.g. OMC-2). While unfavorable due to its chemical properties (see Sect.~\ref{subsec:n2h+}), prominent N$_2$H$^+$ emission is nonetheless detected in the warmer environments of OMC-1 and the Flame Nebula at temperatures $T_\mathrm{K}>$~35~K \citep[e.g.][]{hacar20}. In addition to these global properties, we observe significant temperature variations in those regions traced in N$_2$H$^+$. Several N$_2$H$^+$ emission patches qualitatively show temperature gradients of $\sim5-10$~K within scales of $\sim0.1$~pc from their peak towards their edges. While these variations are typical in regions such as OMC-2, the gradients are more pronounced in regions directly exposed to the feedback from massive stars, such as the Flame Nebula and OMC-1. 

To further explore the correlation between the kinetic temperature and the N$_2$H$^+$ emission, we qualitatively compared the latter with previous temperature estimates at high resolution in some of our regions. Interferometric observations of NH$_3$ (1,1), (2,2) with the Very Large Array (VLA) determined temperature maps at resolutions similar to the one of our ALMA+IRAM-30m maps in OMC-1 \citep[8$''$;][]{wise98} and OMC-2/-3 \citep[5$''$;][]{li13}. Although less extended than our IRAM-30m observations, these maps probe almost the same gas as N$_2$H$^+$ \citep[expected given the chemical connection between the two molecules; e.g.][]{aikawa05}, and show a similar picture as previously depicted: the typical temperature in all three regions is $T_\mathrm{K}\lesssim25$~K, which rises up and above $>40$~K only towards the centre of OMC-1. Since these archival measurements are available only in a sub-sample of our survey, we stick to a fully qualitative comparison. However, and despite the coarser resolution, our IRAM-30m maps not only capture the main features and the temperature ranges seen in the VLA maps, but also extend our knowledge of the gas kinetic temperature further by targeting the material surrounding the N$_2$H$^+$ emission as well.

\subsection{Column density and dense gas properties}\label{subsec:densegasprop}

Beyond the description of the N$_2$H$^+$ intensity variations, we aim to characterise the gas organisation in the survey down to 2000 au. It is therefore mandatory to determine which column densities are effectively sampled in our N$_2$H$^+$ maps. Originally computed as $I$(N$_2$H$^{+}$)/$T_\mathrm{K}$, the temperature-corrected N$_2$H$^+$ integrated intensity ($I_\mathrm{T}$(N$_2$H$^{+}$)) is found to tightly correlate with the \textit{Herschel} total column density N(H$_2$) in OMC-1 and OMC-2 at a resolution of 30$''$ \citep{hacar18}. We relied on a similar calibration to determine the column density maps of our regions by combining the ALMA+IRAM-30m N$_2$H$^+$ maps with the IRAM-30m temperature maps, albeit at different resolutions. As seen in the previous Section, our IRAM-30m observations recover temperature estimates in close agreement with previous studies at a resolution comparable to our N$_2$H$^+$ maps. We therefore assumed, as a first order approximation, $T_\mathrm{K}$ constant within each IRAM-30m beam and associated this temperature pixel by pixel to the ALMA+IRAM-30m N$_2$H$^+$ maps\footnote{We note some small artefacts in our high spatial resolution (4.5$''$) N(H$_2$) maps due to this combination of the high-resolution ALMA+IRAM-30m data (4.5$''$) with the coarser $T_\mathrm{K}$ maps (30$''$).}. 
The correlation between $I_\mathrm{T}$(N$_2$H$^{+}$) and N(H$_2$) that is employed follows \citet{hacar18}, and it is empirically described as
\begin{equation}
    \Bigg(\frac{{\rm{N}}({\rm{H_2}})}{[{\rm{cm}}^{-2}]}\Bigg) = 67.4\times10^{21}~\Bigg(\frac{I_{\rm{T}}({\rm{N_2H^+}})}{[{\rm{km~s}}^{-1}]}\Bigg) ~+~ 12\times10^{21},
    \label{eq:caln2hp}
\end{equation}
with $I_\mathrm{T}$(N$_2$H$^{+}$) as $I$(N$_2$H$^{+}$)/($J(T_\mathrm{K}) - J(T_\mathrm{bg})$), according to \citet{tafalla23}.
The above correlation comes from the expected increase of line excitation as a function of column density, modulated by the temperature (linear term), plus a column density floor value needed to chemically form N$_2$H$^+$ in the gas phase (offset).
Further tests indicate that Eq.~(\ref{eq:caln2hp}) describes the column densities in all other sources of our survey. Adopting Eq.~(\ref{eq:caln2hp}), we obtained individual column density N(H$_2$) maps at high spatial resolution (4.5$''$ or $\sim$~2000~au) for the seven ALMA fields in our sample (Fig.~\ref{fig:NH2}). While predictions for the floor term at cloud scales exist \citep[e.g.][]{tafalla23}, this threshold may change locally at our resolution, thus biasing our N(H$_2$) estimate. We therefore disregard the floor term in Eq.~(\ref{eq:caln2hp}) hereafter, thus arguing in terms of column density of the dense gas only, and all the physical properties derived will be intended as for the dense gas.

Figure~\ref{fig:NH2} shows the column density maps for all seven regions in the survey at 2000 au resolution. Not surprisingly, and thanks to the calibration from Eq.~(\ref{eq:caln2hp}), the observed variations and morphology of our N(H$_2$) maps closely follow those seen in the integrated intensity of the N$_2$H$^+$ (1$-$0) line (Fig.~\ref{fig:In2h+}). An inspection of these new N(H$_2$) maps confirms the presence of a rich substructure comprising narrow, high contrast gas features in all our clouds.

\begin{figure}[tbp]
    \centering
    \includegraphics[width=0.99\linewidth]{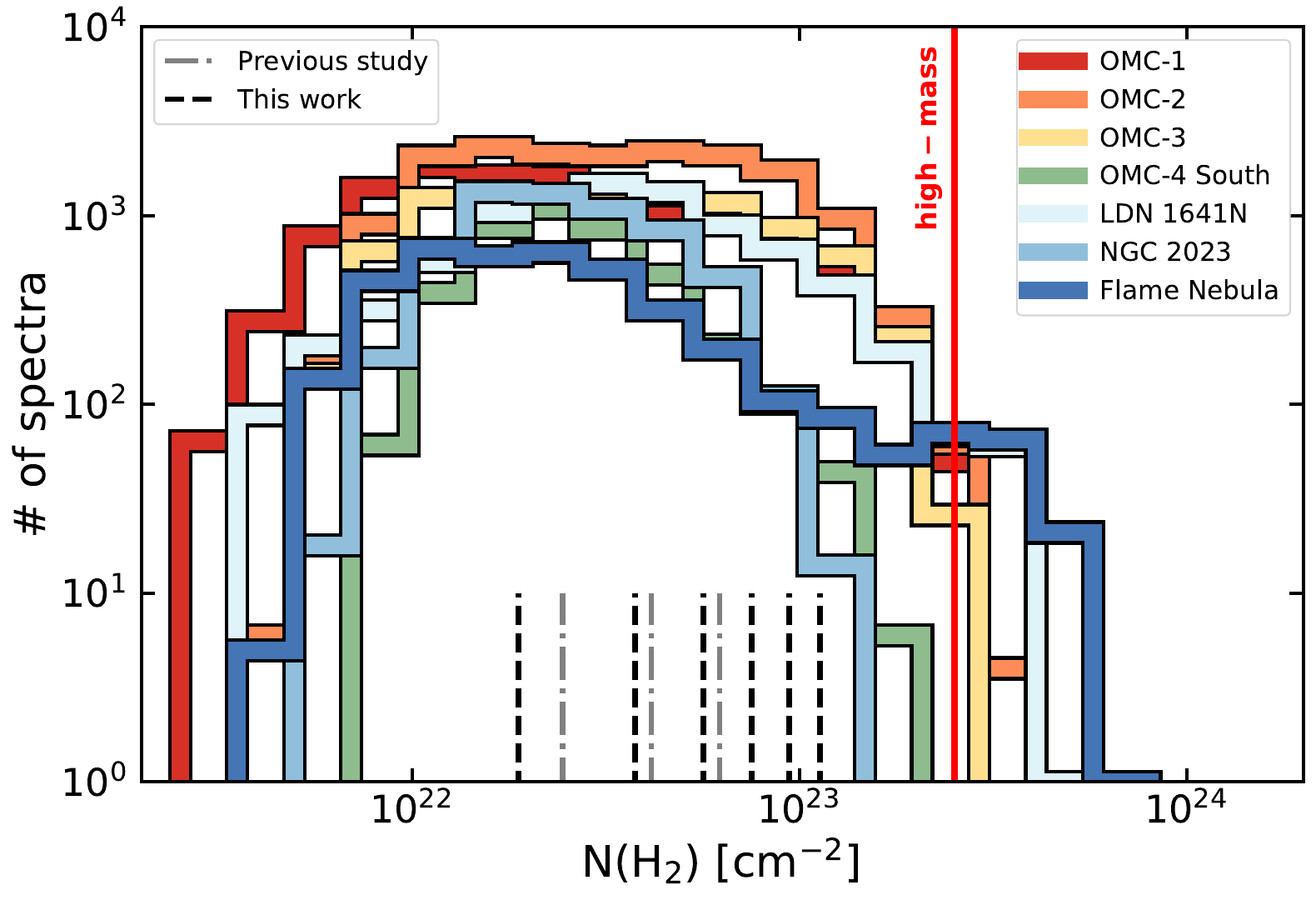}
    \caption{Histogram of the total column density N(H$_2$) detected in N$_2$H$^+$ in the different fields explored by the EMERGE Early ALMA Survey (see colours in legend). Three dashed grey lines (at [26, 44, 66]~A$_\mathrm{V}$) indicate the column density thresholds used by \citet{hacar18} during the analysis of the OMC-1 and OMC-2 regions. The additional six solid black lines (at [20, 40, 60, 80, 100, 120]~A$_\mathrm{V}$) indicate the new HiFIVe thresholds used in this work in order to cover the wide dynamic range of column densities in our entire sample.
    The solid red line shows instead the theoretical prediction for high-mass star-formation \citep{krum08}.}
    \label{fig:cumulN}
\end{figure}

The high dynamic range of our ALMA+IRAM-30m data (see Sect.~\ref{subsec:n2h+}) is further stretched by the correction for $T_\mathrm{K}$.
Figure~\ref{fig:cumulN} shows an histogram per region including all the pixels in our maps with $I\mathrm{(N_2H^+)}\geq$~2.5~K~km~s$^{-1}$, selection criterion later applied in the analysis (see Sect.~\ref{sec:analysis}). When considering these fields, the whole survey spans almost three orders of magnitude in column density, between $\mathrm{N(H_2)}\sim 2\times10^{21}$~cm$^{-2}$ and $\sim10^{24}$~cm$^{-2}$.
The material surveyed by our N$_2$H$^+$ ALMA maps at 2000~au resolution is typically found at column densities N(H$_2$) between 10$^{22}$~cm$^{-2}$ and 10$^{23}$~cm$^{-2}$. These values are one order of magnitude denser, on average, than the peak column densities reported for \textit{Herschel} filaments \citep{arzoumanian19}. Not all clouds exhibit the same range in N(H$_2$), however (see also Paper I). Diffuse regions such as OMC-4 South or NGC~2023 show $\mathrm{N(H_2)}\sim10^{23}$~cm$^{-2}$ at maximum, while regions particularly active in star formation, such as OMC-2 or OMC-3, display column densities above $\mathrm{N(H_2)}>10^{23}$~cm$^{-2}$. Interestingly, the highest values in the cumulative distributions are found exceeding the theoretical threshold for high-mass star formation \citep[N$(\mathrm{H_2})\gtrsim10^{23.4}$~cm$^{-2}$;][]{krum08} towards OMC-1 and the Flame Nebula, both hosts of O-type stars.

As selective tracer of densities above $n\mathrm{(H_2)}\gtrsim10^5$~cm$^{-3}$, N$_2$H$^+$ samples high column density (N$(\mathrm{H_2})\gtrsim10^{22}$~cm$^{-2}$) and cold ($T_\mathrm{K}=10-25$~K) material directly connected to the formation of young protostars in our fields. 
When observed at 2000~au resolution, this gas appears to be highly structured forming complex networks of narrow and elongated filaments at sub-parsec scales. Independently of the evolutionary stage (young vs evolved) and stellar activity (low- vs high-mass), our results suggest a common organisation of the dense gas prior to the formation of stars.
The enhanced resolution and dynamic range of our ALMA+IRAM-30m maps makes the EMERGE Early ALMA Survey an ideal sample of regions to investigate the initial conditions for star formation across a wide range of physical properties.

\begin{figure*}[tbp]
\centering
\includegraphics[width=0.99\linewidth]{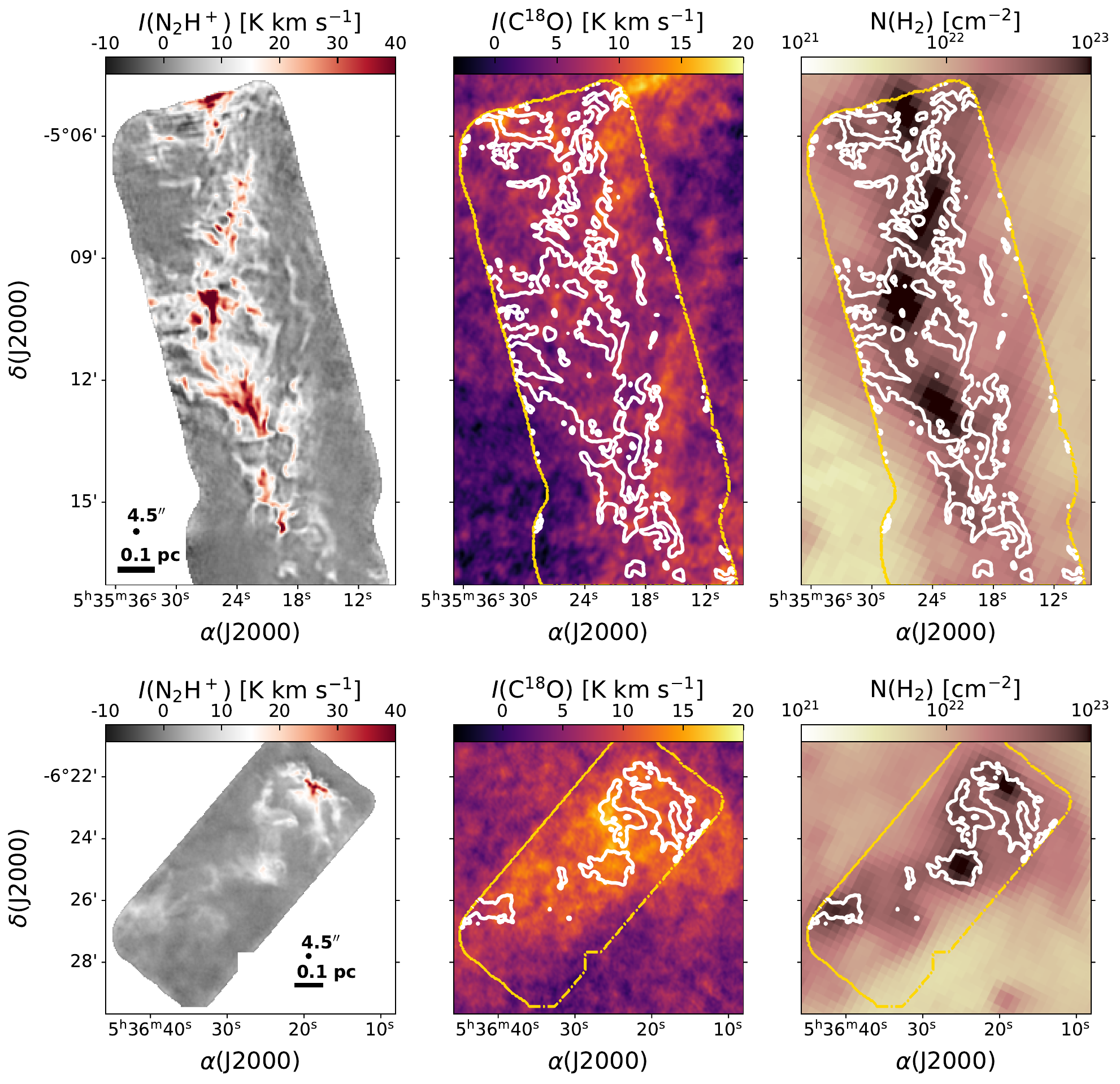}
\caption{OMC-2 (upper panels) and LDN~1641N (lower panels) as seen in different surveys. (\textbf{Left panels}): ALMA+IRAM-30m integrated emission of N$_2$H$^+$ (1$-$0) at 4.5$''$. (\textbf{Central panels}): CARMA-NRO integrated emission of C$^{18}$O (1$-$0) at 8$''$ \citep{suri19}. The white dotted line is the footprint of our ALMA+IRAM-30m observations. (\textbf{Right panels}): \textit{Herschel}+\textit{Spitzer} dust continuum at 36.2$''$ \citep{lombardi14}. The gold dotted line is the footprint of our ALMA+IRAM-30m observations, while the white contours represent $I(\mathrm{N_2H^+})>5$~K~km~s$^{-1}$, twice as those in Fig.~\ref{fig:In2h+} to highlight the tight correspondence with the high column density material.}
\label{fig:surveycompar}
\end{figure*}

\subsection{Comparison with previous surveys: parsec-scale depletion}\label{subsec:comparsurv}

The Orion complex is a prime choice in star-formation studies including a wide number of ancillary observations at all wavelengths and scales. 
Due to their large angular scales, parsec-scale surveys of the Orion A and B clouds were typically restricted to dust continuum \citep{john99,lombardi14,konyves20} or low-J CO \citep{bally08,Shimajiri2014,Nishimura2015,ork19,yun21} observations at low angular resolutions ($\gtrsim15$\arcsec). Only in recent times similar continuum \citep{kainul17,schuller21} and CO \citep{Kong2018,suri19} observations at intermediate resolutions (8\arcsec) covered larger areas of these clouds. In contrast, high-resolution observations ($<5$\arcsec) of high-density tracers were so far restricted to small fields containing specific targets with bright emission lines \citep[e.g.][]{wise98,shima23}.

Tracers such as CO efficiently probe the diffuse, low-density gas in clouds, but cannot study the properties of the dense, star-forming gas within filaments. 
We demonstrated these differences using the previous CARMA-NRO Orion Survey \citep{Kong2018} as equivalent survey in terms of resolution (8\arcsec) and covering most of our targets (OMC-1/2/3/4 South and LDN~1641N) with different CO isotopologues. Figure~\ref{fig:surveycompar} shows the integrated-intensity emission of N$_2$H$^+$ (1$-$0) obtained by our ALMA+IRAM-30m maps (left panel, 4.5$''$ or $\sim2000$~au; this work) paired with the C$^{18}$O (1$-$0) emission observed in the CARMA-NRO survey \citep[central panel, 8\arcsec, or $\sim3000$~au;][]{suri19} in OMC-2 (upper panel) and LDN~1641N (lower panel). These two high-resolution maps can be compared against the total column density map derived from dust continuum emission observed by \textit{Herschel} at low resolution \citep[right panel, 36$''$ or $\sim0.07$~pc;][]{lombardi14}. 
Despite the different resolution, the morphology and peaks of the N$_2$H$^+$ emission detected in our ALMA maps closely follow the shape and distribution of the high-column density material above N$(\mathrm{H_2})\gtrsim10^{22}$ cm$^{-2}$ detected in continuum. 
The C$^{18}$O emission shows instead a much flatter distribution with no clear correspondence with the total gas column density in neither OMC-2 (upper panels) nor LDN~1641N (lower panels) above a few $10^{21}$~cm$^{-2}$. These differences in emission are likely the consequence of severe CO depletion operating at parsec-scales in our targets (see Sect.~\ref{sec:observations}). The freeze-out of CO could therefore hamper its use as tracer of the gas distribution within star-forming regions such as OMC-2, preventing any direct comparison with dense tracers such as N$_2$H$^+$. In contrast, Fig.~\ref{fig:surveycompar} highlights N$_2$H$^+$ as a unique tracer for the dense filamentary sub-structure within the EMERGE sample.

\section{Systematic search for velocity-coherent structures}\label{sec:analysis}

The EMERGE Early ALMA Survey employs seven high-resolution maps of regions in Orion for the analysis. These maps show a complex morphology of the gas, which is described by a total of $\sim170,000$ spectra. To extract quantitative information from this data in a systematic way, a fully, or at least partly, automated analysis is required. This work employs a semi-automatic analysis carried out in four major steps: we first performed a supervised fitting of the spectra through customised GILDAS/CLASS routines \citep{pety05,gil13} to retrieve the main spectral line parameters (centroid velocity, linewidth, and peak intensity). We then applied a selection on the spectra, to ensure their quality and signal-to-noise (S/N), based on several criteria (e.g. the radial velocity of the source). Third, we ran the automatic identification of velocity-coherent structures throughout different column density thresholds by applying the Hierarchical Friends-In-Velocity (HiFIVe) algorithm \citep{hacar18} on the selected fields. We finally ran the automatic fitting of the column density radial profiles on all the identified structures with the publicly available routine FilChap \citep{suri19}.

Among the total of $\sim170,000$ N$_2$H$^+$ spectra sampled in our EMERGE Early ALMA survey, we effectively fitted $\sim55,000$ spectra. From these spectra our selection extracted $\sim45,000$ high-quality fits comprising at least one component with $\mathrm{S/N}\geq3$.
One third of the selected spectra ($\sim17,000$) has multiple components fulfilling the selection criteria, leading to a total of $\sim57,000$ independent velocity components available for the analysis (see Table \ref{tab:gen_prop}). In this Paper III, we discuss the results of HiFIVe in the identification and characterisation of fibers alone. The results derived from the radial profile fitting, and the corresponding discussion, are left for another paper of this series \citep[][hereafter Paper IV]{socci24}. We refer the reader to the Appendices~\ref{sec:filchap}-\ref{sec:parexpl} for a thorough discussion on the algorithm structure, the selection criteria and the comparison with previous studies in OMC-1 and OMC-2. Here we only report the conclusions drawn at the end of these Appendices: first, we derive physical properties for the structures in OMC-1, OMC-2 consistent with those of \citet{hacar18}, both by applying their original setup (e.g. same column density thresholds, linking length) and different ones (see Appendix~\ref{subsec:comparhacar18}). Second, different setups applied to all regions composing our survey produce results consistent one with the other and with the final setup chosen in our analysis (see Appendix~\ref{sec:parexpl}). Ultimately, the algorithmic choices, condensed in the analysis setup, have only a minor impact on the physical results discussed throughout the paper.

\begin{figure*}[tbp]
    \centering
    \includegraphics[width=\linewidth]{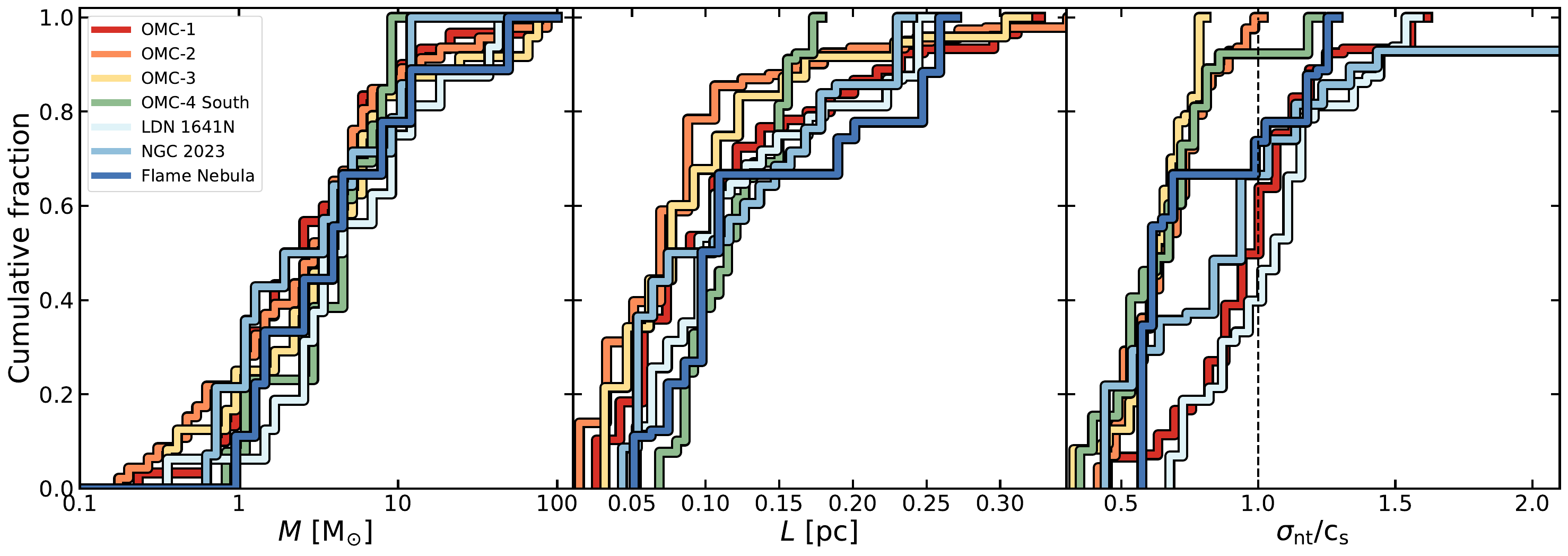}
    \caption{Cumulative distributions of the main physical properties obtained for the velocity-coherent structures identified in our EMERGE Early ALMA sample (see colours in legend). (\textbf{Left panel}) Total mass of dense gas; (\textbf{Central panel}) Total length; and (\textbf{Right panel}) Non-thermal velocity dispersion in units of the sound speed $\sigma_\mathrm{nt}/c_\mathrm{s}$. A dashed black line shows the transition between sub- and trans-sonic regime (i.e. $\sigma_\mathrm{nt}/c_\mathrm{s}=1$).}
    \label{fig:first_results}
\end{figure*}

\begin{table*}[tbp]
    \caption{General properties obtained for the dense gas component from the analysis.}
    \centering
    \small
    \begin{tabular*}{\linewidth}{c@{\hspace{1\tabcolsep}}c@{\hspace{1\tabcolsep}}c@{\hspace{1\tabcolsep}}c@{\hspace{1\tabcolsep}}c@{\hspace{2\tabcolsep}}c@{\hspace{2\tabcolsep}}c@{\hspace{2\tabcolsep}}c@{\hspace{2\tabcolsep}}c@{\hspace{2\tabcolsep}}c@{\hspace{2\tabcolsep}}c@{\hspace{2\tabcolsep}}c@{\hspace{1\tabcolsep}}c}
    \hline
    Source & \multicolumn{2}{c}{Spectra} & Recovered & Fibers & $M_{\rm{tot}}$ & $\sigma_\mathrm{nt}/c_\mathrm{s}$ & $L$ & \textit{m} & \textit{m}/\textit{m$_\mathrm{vir}$} & $\nabla T_{\mathrm{K}}$ & $N_0$ & $FWHM$ \\
     & Fitted & \# comps. & [\%] & [\#] & [M$_{\odot}$] & [] & [pc] & [M$_{\odot}$ pc$^{-1}$] & [] & [K pc$^{-1}$] & [$10^{22}$ cm$^{-2}$] & [pc] \\
    \hline \\ [-2ex]
    OMC-1 & 9562 & 12442 & 96 & 30 & 209 & 0.99 & 0.094 & 26 & 0.33 & 64 & 4.9 & 0.034$^{+0.010}_{-0.007}$ \\ [0.7ex]
    OMC-2 & 10231 & 11744 & 98 & 46 & 350 & 0.66 & 0.076 & 40 & 0.90 & 76 & 6.7 & 0.043$^{+0.018}_{-0.009}$ \\ [0.7ex]
    OMC-3 & 8169 & 11786 & 98 & 24 & 257 & 0.64 & 0.085 & 50 & 1.08 & 75 & 6.4 & 0.050$^{+0.022}_{-0.010}$ \\ [0.7ex]
    OMC-4 South & 8284 & 4715 & 96 & 13 & 64 & 0.66 & 0.123 & 39 & 0.96 & 96 & 3.0 & 0.059$^{+0.019}_{-0.011}$ \\ [0.7ex]
    LDN~1641N & 7414 & 8280 & 99 & 16 & 168 & 1.09 & 0.105 & 51 & 0.97 & 150 & 4.1 & 0.063$^{+0.026}_{-0.012}$ \\ [0.7ex]
    NGC~2023 & 7614 & 4904 & 92 & 14 & 66 & 0.94 & 0.095 & 25 & 0.70 & 75 & 3.7 & 0.074$^{+0.025}_{-0.013}$ \\ [0.7ex]
    Flame Nebula & 4045 & 3714 & 95 & 9 & 94 & 0.63 & 0.111 & 27 & 0.85 & 110 & 4.2 & 0.038$^{+0.014}_{-0.013}$ \\ [0.5ex]
    \hline \\ [-2ex]
    Sample & 55319 & 57585 & 96 & 152 & 1209 & 0.74 & 0.091 & 36 & 0.78 & 80 & 5.1 & 0.05$^{+0.02}_{-0.01}$ \\ [0.5ex]
    \hline
    \end{tabular*}
    \label{tab:gen_prop}
\end{table*}

\section{Dense fiber networks in Orion}\label{sec:fibersurvey}

We identified 152 velocity-coherent structures in the EMERGE Early ALMA Survey by applying HiFIVe on the selected spectra (Sect.~\ref{sec:analysis}). Our selection criteria and further analysis recover $\gtrsim95\%$ of the total velocity components extracted from our spectra across the survey (Table~\ref{tab:gen_prop}). 
The total number of structures identified in this work (152) increases by a factor of 3 the number of velocity-coherent structures determined in OMC-1 and OMC-2 (55). The larger statistics on which our survey leverages allows the exploration of a wider dynamic range of physical properties and environmental effects.
In the following, we discuss the statistical properties of these velocity-coherent structures, their nature as fibers, and their connection to star formation.
A table with the physical properties derived per fiber is found in Appendix~\ref{sec:mapsandtab} (online version available through Zenodo), along with the finding charts for the position of each fiber (see Fig.~\ref{fig:numberedfib}).

\subsection{Basic properties: mass and length distributions}\label{subsec:MandL}

Mass and length define the two most basic physical properties of the gas structures derived from our observations. We estimated the total mass $M$ of each structure identified with HiFIVe by adding the mass contribution of all its constituents gas components (N$_{\rm{comp}}$) as follows:
\begin{equation}
    M = \sum_i^{{\rm{N_{comp}}}}~{\rm{N}}_i({\rm{H_2}})~A_{\rm{pix}}~\mu_{\rm{H_2}}~m_{\rm{p}}.
    \label{totmass}
\end{equation}
In the above equation, N$(\mathrm{H_2})$ is the column density per point, estimated using Eq.~(\ref{eq:caln2hp}), $\mu_\mathrm{H_2}$ is the molecular mass of H$_2$ \citep[$=2.81$; see][]{kauff08,aspl21}, $m_\mathrm{p}$ is the proton mass, and $A_\mathrm{pix}=\bigg(D~\frac{2\pi~pxl}{3600\times360}\bigg)^2$ the pixel area, with $pxl=2.25$\arcsec, pixel size of our Nyquist sampled maps, and $D$ the corresponding distance to the region of interest (see Table \ref{tab:sample_prop}).

Figure~\ref{fig:first_results} (left panel) displays the cumulative distribution of masses in each of our EMERGE targets. The identified structures range approximately three orders of magnitude in mass, within $\sim0.1-100$~$\rm{M}_\odot$. The median mass per structure falls within $\sim2.5-5~\mathrm{M}_{\odot}$ in our survey, consistently with the majority of the structures ($\sim80\%$) showing masses below $<10~\mathrm{M}_{\odot}$. Nonetheless, several structures exceed instead masses of $>20~\mathrm{M}_{\odot}$ in active star-forming regions such as OMC-1 and OMC-2 (see a further discussion in Sect.~\ref{subsec:linemasses}).
Despite comprising high- to low-mass star-forming regions (e.g. OMC-1 vs NGC~2023), our targets do not show any systematic difference in their mass distributions, which are instead all continuous within the aforementioned dynamic range.

In spite of their similar mass distribution, the frequency of structures per region widely varies in our sample. Regions such as OMC-1 and OMC-2 contain more than 30 velocity-coherent structures each; others such as OMC-4~South and the Flame Nebula only contribute with $\sim10$ per region. The amount of objects translates into noticeable differences in the total mass of dense gas ($M_\mathrm{tot}$) recovered per target (see Table \ref{tab:gen_prop}), with OMC-1, OMC-2 and OMC-3 being the most massive regions ($>200~\mathrm{M}_{\odot}$). We recovered a total amount of dense gas of 1209~M$_{\odot}$ for the whole survey at the ALMA resolution (see Table \ref{tab:gen_prop}). This total mass corresponds to $\sim 70\%$ of the estimated $\sim1700$~M$_{\odot}$ dense gas mass surveyed in Paper I at the IRAM-30m resolution within an area of $\sim1.5 \times 1.5$~pc$^2$ around our targets. Even if less extended in size (i.e. $\sim 0.3-0.4$~pc$^2$), our ALMA+IRAM-30m maps still capture the majority of the dense gas within these regions\footnote{As discussed in Paper I, our N$_2$H$^+$ (1$-$0) ALMA+IRAM-30m maps recover $\gtrsim90\%$ of the emission detected in the single-dish data within the footprint of our ALMA mosaics. We remark here that the reported differences in mass are therefore only due to the different coverage of the single-dish and ALMA maps and not due to sensitivity issues or interferometric filtering effects.}. 

We characterised the length of the velocity-coherent structures in our sample measured along their main spine as determined by HiFIVe (see Sect.~\ref{subsec:identification}). We display these axes for all structures identified in our sample in Fig.~\ref{fig:NH2} (red segments), plotted over the corresponding total gas column density map of each region. The distribution and orientation of these axes follow those same elongated features seen in the integrated intensity of N$_2$H$^+$ (Fig.~\ref{fig:In2h+}) and gas column density (Fig.~\ref{fig:NH2}). 
Recent simulations showed the potential risks of interpreting some of the velocity features identified in the PPV space as gas structures \citep{zamora17,clarke18}. Nonetheless, the good agreement between the identified axes and actual region features suggests a close correspondence between the velocity-coherent structures identified in our PPV analysis and the true mass distribution of the dense gas. 

We further explore distribution of lengths in our sample by inspecting their cumulative distribution per region in Fig.~\ref{fig:first_results} (central panel). Our velocity-coherent structures show lengths varying an order of magnitude within $\sim0.03-0.3$~pc. More than half of these structures has a length within $\sim0.05-0.2$~pc with a corresponding median value for the whole sample of 0.09~pc. These structures are therefore well sampled in our analysis by more than 5 beams, 10 on average, at the resolution of our maps ($\theta_\mathrm{beam}=$~4.5\arcsec or $\sim0.009$~pc). 
Only $\sim3\%$ of the structures is instead unresolved in our survey by showing lengths below 0.027~pc (i.e. $< 3\times\theta_\mathrm{beam}$).
Slight variations of the length distributions are seen in the survey, with OMC-4 South showing the shortest dynamic range (i.e. the steepest cumulative distribution). Aside from this last exception, the other regions show similar length distributions and all exhibit a certain degree of skewness with several structures having lengths above $\gtrsim0.2$~pc.

We computed the aspect ratios of the identified structures as $AR = L/FWHM$, where $FWHM$ is the full width at half maximum of the column density radial profile fitted by FilChap (see Paper IV for the derivation). We remark here that HiFIVe identifies structures only through their velocity-coherence in the PPV-space (see Sect.~\ref{subsec:identification}). Thus, when determining their axis, the aspect ratio ($AR$) of these objects remains unconstrained and it is not limited to $AR\leq3$, usually employed to define filaments \citep[e.g.][]{arzoumanian19}. The proportion between objects with $AR < 3$ and $AR\geq3$ in our sample is then 70:30 with a corresponding median value of $\langle AR\rangle = 2.0^{+1.1}_{-0.7}$, which suggests mildly elongated objects to be dominant throughout the survey. 
This proportion finds its extremes in NGC~2023 with up to $\sim90\%$ of the structures having $AR\leq3$ and in Flame Nebula with $\sim55\%$ of them having $AR>3$. The combination of different geometries within single regions, and in the survey at large, is somewhat expected given the complex organisation of the dense gas (see Sect.~\ref{sec:densegas}). By inspecting the lengths recovered throughout hierarchies in our analysis, we identify a decrease in $AR$ for increasing column density thresholds (or hierarchy) overall. These findings suggest a potential transition from filamentary (prolate) to round (oblate) geometries towards higher column densities, a topic which will be further investigated in other papers of this series. A further discussion on the aspect ratios, their comparison with the results of \citet{hacar18}, and their impact when comparing to theoretical models will be addressed in Paper IV.

\begin{figure}[tbp]
    \centering
    \includegraphics[width=\linewidth]{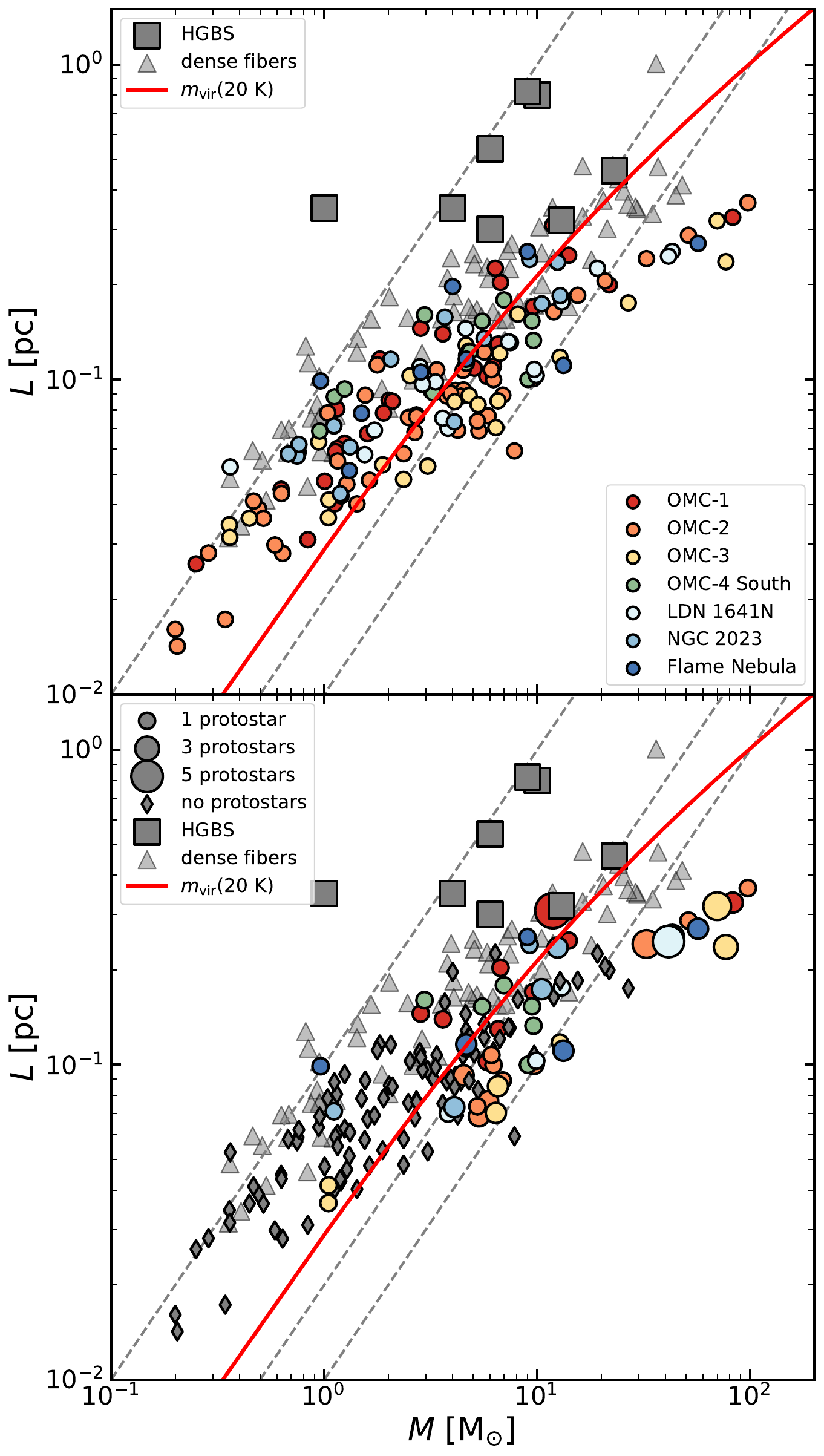}
    \caption{Comparison between the mass and length of the velocity-coherent structures identified in our survey. {\bf (Upper panel):} Mass and length values of the 152 velocity-coherent fiber-like structures identified in this work (colour-coded circles; see legend). Additional points indicate the mean properties of the parsec-size filaments explored by the HGBS in nearby clouds \citep[grey squares;][]{arzoumanian19} as well as the previously reported fibers \citep[grey triangles; see][and references therein]{hacar22}. 
    {\bf (Lower panel):} Same as upper panel but showing those star-forming fibers in our sample highlighted in colour and shown with symbols sizes proportional to the number of protostars in them (see legend).
    In both panels, a curve (solid red line) indicates the virial line mass $m_\mathrm{vir}$ computed at 20~K (see Sect.~\ref{subsec:linemasses}). Also, diagonal lines (dashed grey lines) represent constant line masses of 10, 50, and 100~M$_{\odot}$~pc$^{-1}$, respectively.}
    \label{fig:mlpaper}
\end{figure}

Due to the hierarchical nature of the ISM, filaments are reported at all scales in our Galaxy depending on the tracer, detection technique, and resolution used for their study. As discussed in Sect.~\ref{sec:introduction}, this empirically-driven approach lead to a classification of filaments into families, which range from the giant filaments ($\sim100$~pc in size) through the nearby filaments ($\sim1-10$~pc scales), and down to the fibers ($\sim1000$~au regime). Each one of these families exhibits a characteristic mass range, and therefore, occupies a distinct area in the Mass-Length (M-L) parameter space for filaments \citep[see][for a review]{hacar22}. 

To put our results in the context of filament families, we compared the masses and lengths of our velocity-coherent structures with those of other filamentary structures from the literature.
Figure~\ref{fig:mlpaper} (top panel) displays the masses and lengths extracted from our EMERGE Early ALMA Sample (colour-coded points; see legend) paired with ancillary estimates. We first included the mean filament properties sampled by the {\it Herschel} Gould Belt Survey (HGBS) in different nearby clouds, including Orion B \citep[grey squares;][]{arzoumanian19}. We then added the same mean properties reported for fibers in clouds like Perseus, Serpens, and Orion from other studies based on the N$_2$H$^+$ emission \citep[grey triangles\footnote{We corrected the masses of the fibers identified in \citet{hacar17} and \citet{hacar18} by a factor of 2.81/2.33 to account for the different $\mu_\mathrm{H_2}$ used in these studies.}; see][and references therein]{hacar22}. From the comparison, the velocity-coherent structures in our survey cover a similar, if not broader, dynamic range in mass than the previous estimates in local clouds. Thanks to the enhanced resolution of our ALMA observations, we sampled filaments with lengths up to an order magnitude smaller compared to those filaments by {\it Herschel} on average. The reduced size of our structures naturally leads to higher average densities for a similar range of masses, a result expected given the density-selective nature of N$_2$H$^+$ as a tracer (see Sect.~\ref{sec:densegas}). Not surprisingly, the M-L distribution of our velocity-coherent structures shows a significant overlap with the parameter space occupied by the filament family of fibers. Although the distinction between families is largely artificial and primarily determined by observational biases \citep[see][for a discussion]{hacar22}, the location of our structures in the M-L parameter space suggests them to possess properties similar to these small-scale and dense filaments.  

\subsection{Networks of sonic-like fibers}\label{subsec:submot}

The kinematics of the gas is usually evaluated through its total velocity dispersion $\sigma_\mathrm{tot}$. This total velocity dispersion results from the combination of a thermal and a non-thermal component as $\sigma_\mathrm{tot}^2 = \sigma_\mathrm{th}^2 + \sigma_\mathrm{nt}^2$. The relevance of the non-thermal component is usually evaluated in comparison to the gas sound speed ($c_\mathrm{s}(T_\mathrm{K})$), and classified as sonic ($\sigma_\mathrm{nt}/c_\mathrm{s}\leq 1$), tran-sonic ($1< \sigma_\mathrm{nt}/ c_\mathrm{s}<2$), or super-sonic ($\sigma_\mathrm{nt}/c_\mathrm{s} > 2$). Classically, there are multiple approaches to quantify the non-thermal motions within a cloud \citep[e.g.,][]{miesch94}. These non-thermal motions include ordered and turbulent modes, which can both possibly act against the gravitational collapse of the cloud.
For each of the gas components identified in our spectra, the unresolved motions along the line-of-sight can be described by the total velocity dispersion $\sigma_\mathrm{los}$ ($=\Delta v/\sqrt{8~\mathrm{ln} 2}$), from which we can readily determine the non-thermal component as follows:
\begin{equation}\label{eq:non-therm}
    \sigma_{\rm{nt}} = \sqrt{\sigma_{\rm{los}}^2 - \frac{k_{\rm{B}}T_{\rm{K}}}{\mu_{\rm{N_2H^+}}m_{\rm{p}}}},
\end{equation}
where $\sigma_\mathrm{th}=\sqrt{k_{\rm{B}}T_{\rm{K}}/\mu_{\rm{N_2H^+}}m_{\rm{p}}}$ is the thermal broadening of the N$_2$H$^+$ line (with $\mu_{\rm{N_2H^+}}=29$), and where $T_\mathrm{K}$ is taken from our temperature maps (Fig.~\ref{fig:Tk}). The total non-thermal component of the gas accounts for two additional motions, those in the plane of the sky, which we cannot probe with our spectra. Since we have no direct measure of these motions and to facilitate the comparison with previous studies \citep[e.g. the non-thermal motions in][]{hacar18}, we stick hereafter to Eq.~(\ref{eq:non-therm}), acknowledging that the 3D velocity dispersion is likely higher than the one presented \citep[e.g.][]{bertoldi92}. 

We evaluated the non-thermal motions of a single velocity-coherent structure as median of all the non-thermal velocity dispersion components associated to that specific structure by HiFIVe. The non-thermal motions within a structure, however, may be also determined as dispersion of the centroid velocities ($\sigma(V_\mathrm{lsr})$) of its constituents. We therefore computed this estimate per structure in our survey and compared them to $\sigma_\mathrm{nt}$. The two estimates are comparable within a factor of $\sim2-3$ in the range $\sim0.1-0.4$~km~s$^{-1}$ throughout the survey (whether the $\sigma(V_\mathrm{lsr})$ is estimated as dispersion of centroid velocities, $\langle\sigma(V_\mathrm{lsr})\rangle$, or through the velocity gradients as $\nabla V\times FWHM$ or $\nabla V_\mathrm{x}\times L$)\footnote{We considered for both estimates a distribution of inclination angles centred around $\alpha= 45^{\circ}$, for which the projection factor is $\mathrm{tan}~\alpha = 1$.}. We thus consider $\sigma_\mathrm{nt}$ as the parameter describing the non-thermal motions in our fibers hereafter. We assume that $\sigma_\mathrm{nt}$ probes the unresolved, turbulent, and random non-thermal motions along the line of sight, and that these motions act as pressure support against the fiber gravity.

We display the median $\sigma_\mathrm{nt}$ per fiber, in units of the sounds speed ($\sigma_\mathrm{nt}/c_\mathrm{s}$), in Figure~\ref{fig:first_results} (right panel). The majority of our structures shows non-thermal motions that are tran-sonic at most (i.e. $\sigma_\mathrm{nt}/c_\mathrm{s}\leq 2$). In addition, more than half of these structures have sonic, or close to sonic, non-thermal motions (i.e. $\sigma_\mathrm{nt}/c_\mathrm{s}\lesssim 1$; see Table \ref{tab:gen_prop}). Interestingly, we observe this behaviour in low- (e.g. NGC~2023) and high-mass (OMC-1 and Flame Nebula) star-forming regions alike. Only minor differences are seen in the regions composing our sample, such as OMC-2 and OMC-3 where the contribution from sub-sonic structures grows almost to 100\%. Previous results in literature suggested the presence of super-sonic turbulent motions in high-mass star-forming regions \citep[e.g.][]{tan14}. Consistently with previous studies based on the N$_2$H$^+$  emission \citep{lee14,henshaw14,hacar17,sokolov19}, our results demonstrate that the dynamic state of the dense gas is mostly (sub-)sonic, regardless of the star-formation regime and environmental effects \citep[see also][]{hacar18}.

As a unique characteristic among the different filament families, fibers are identified as velocity-coherent, sonic-like structures irrespective of their size \citep{hacar11}.
Fibers are therefore the first sonic-like structures formed out of the turbulent cascade dominating the gas dynamics at larger scales, for which they are suggested as the preferred mode for the gas organisation at high densities. The (tran-)sonic non-thermal motions discussed above for all our velocity-coherent structures (Fig.~\ref{fig:first_results}, right panel), paired with their mass and length distributions, reinforce the classification of these objects as fibers. 
Given the properties similar to those identified in the literature, we hereafter refer to our velocity-coherent structures simply as fibers.

Within the EMERGE Early ALMA Survey, we identified a total of 152 velocity-coherent fibers. As seen from their distribution within our maps (e.g. OMC-2, Fig.~\ref{fig:NH2}), these fibers are organised in networks of different complexity in the seven star-forming regions explored by our survey in Orion. These fiber networks are constituted by distributed groups of $\lesssim 15$ fibers in regions such as NGC~2023, or by densely packed associations of $\gtrsim40$ fibers in regions, as in OMC-2. The morphology of these networks ranges from hub configurations in regions such as OMC-1 (see also Sect.~\ref{subsec:n2h+}), where most fibers have their axes convergent towards the centre of gravity, to regions such as OMC-3 where the fibers mostly align with the parsec-scale structure of the region. In the following sections we investigate how the different network complexity, in particular its surface density of fibers, influences the star-formation regime in each region.

\subsection{Line mass, virialisation, and star formation}\label{subsec:linemasses}

\begin{figure}[tbp]
    \centering
    \includegraphics[width=\linewidth]{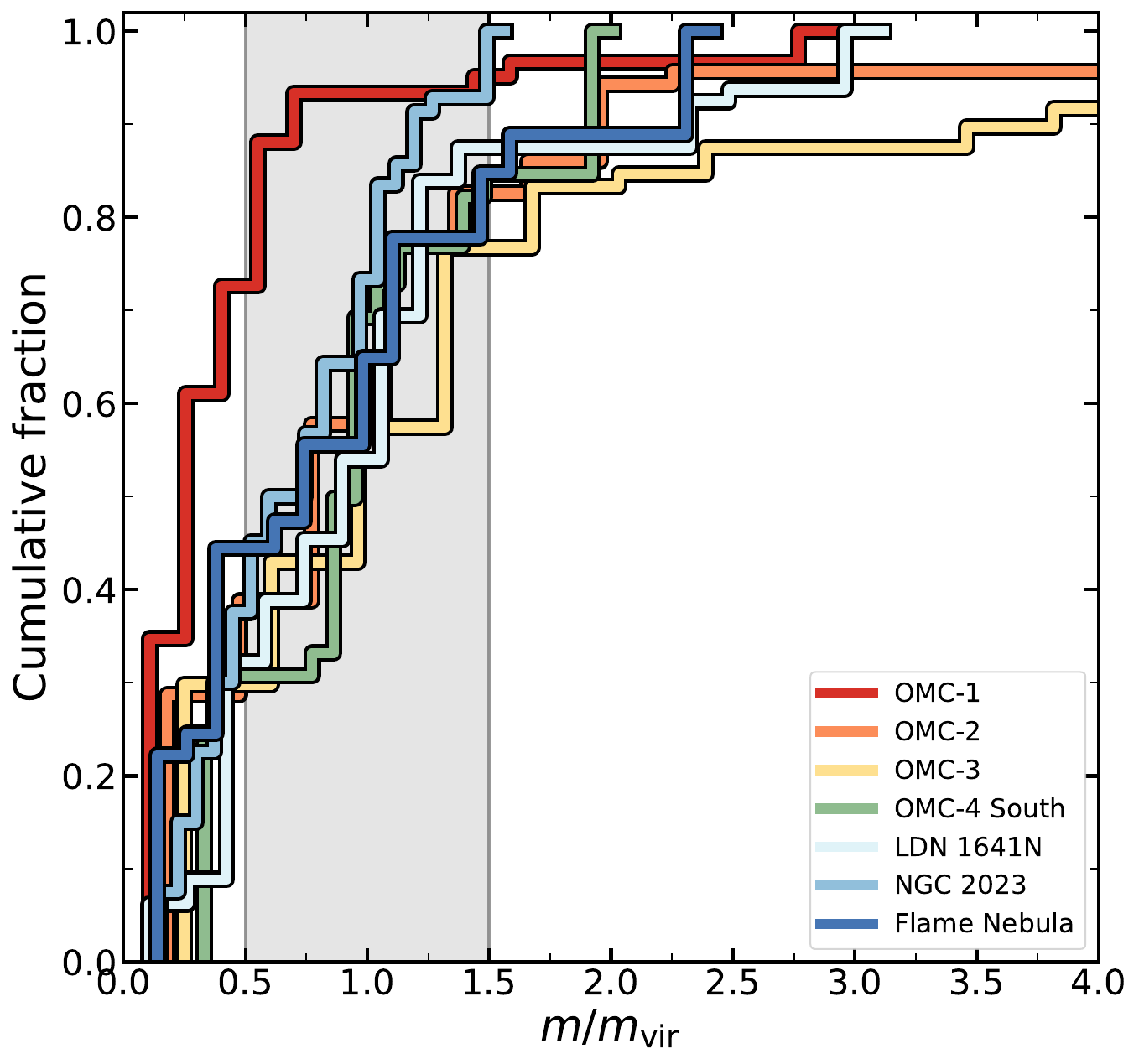}
    \caption{Cumulative distribution of the line masses, in units of the virial mass, for the EMERGE Early ALMA Survey. The black dashed line represents the super-critical limit \citep{hacar22}. The grey shaded area corresponds to virial ratios within $0.5 < m/m_\mathrm{vir} < 1.5$ \citep[see][for a discussion]{inut97}.}
    \label{fig:gen_prop_plot}
\end{figure}

The ratio between the mass and length in filaments and fibers (i.e. the line mass; \textit{m}) is classically employed to evaluate their stability against radial collapse.
This line mass is usually compared against the expected critical mass of an isothermal, infinite filament in hydrostatic equilibrium \citep[$m_\mathrm{crit}=2c_\mathrm{s}^2/G$;][]{ostriker64}. The critical line mass has a similar role as the Jeans mass for molecular clouds (e.g. \citealp{klessen00}). Only sub-critical filaments (with $m/m_\mathrm{crit}\leq1$) are able to sustain themselves against their self-gravity and find stability through an external pressure; super-critical filaments (with $m/m_\mathrm{crit}>1$), on the other hand, are expected to experience a fast collapse under their own gravity.
In this idealised scenario, super-critical filaments are expected to collapse in timescales comparable to their free-fall time. These timescales are faster than any fragmentation process within filaments, which would therefore become unable to host cores and stars \citep{inut97}.

In the absence of magnetic fields, the notion of critical mass may be extended to include the contribution from non-thermal motions (see Sect.~\ref{subsec:submot}) into the virial line mass
\begin{equation}
    m_{\rm{vir}} = \frac{2(c_{\rm{s}}(T_{\rm{K}})^2 + \sigma_{\rm{nt}}^2)}{G}=\frac{2\sigma_{\rm{tot}}^2}{G}.
    \label{mvirial}
\end{equation}
Similarly to $m_\mathrm{crit}$, the virial mass $m_{\rm{vir}}$ is usually interpreted as a stability criterion as well. The non-thermal motions $\sigma_{\rm{nt}}$ are therefore assumed to contribute as an additional (isotropic) pressure term included in the total velocity dispersion $\sigma_\mathrm{tot}$. The virial mass already implies the assumption of isotropic motions, therefore we should include a factor of 3 in the above equation as the sound speed and the non-thermal velocity dispersion measured along the line of sight (see Sect.~\ref{subsec:submot}) are equal to their corresponding components in the plane of the sky \citep{bertoldi92}. We instead kept the estimate given by Eq.~(\ref{mvirial}) to better compare with the scaling relation in Fig.~\ref{fig:mlpaper} which was determined from the 1D velocity dispersion of filaments across the Galaxy \citep[see][for the definition]{hacar22}\footnote{We remark here that, while changing the absolute values of the $m/m_\mathrm{vir}$ ratios, the presence of a factor of 3 does not change the conclusions derived throughout the Section.}. In opposition to an oversimplified picture of a stability criterion, a critical condition with respect to $m_{\rm{vir}}$ should be instead interpreted as energy equipartition in the radial direction: virialised filaments (with $m/m_{\rm{vir}}\sim 1$) are likely dynamic structures which evolve via multiple processes, such as fragmentation, accretion, and/or collapse \citep[see][for a discussion]{hacar22}. 

We calculated the virial ratio $m/m_{\mathrm{vir}}$ for each of the fibers identified in our survey. The sound speed ($c_{\rm{s}}(T_\mathrm{K})$) and non-thermal motions ($\sigma_{\mathrm{nt}}$) for these fibers are estimated from their median gas temperature and the median velocity dispersion of their single components (see Sect.~\ref{subsec:submot}), respectively.
We display the cumulative distribution of $m/m_{\mathrm{vir}}$ for each of the regions in our EMERGE Early ALMA Survey in Fig.~\ref{fig:gen_prop_plot}. In most regions, the observed $m/m_{\mathrm{vir}}$ ratios of our fibers follow similar distributions: almost half of the objects ($\sim40\%$) shows $m/m_\mathrm{vir} < 0.5$ throughout the sample; another $\sim45\%$ appears instead to be virialised within $0.5 < m/m_\mathrm{vir} < 1.5$ (shaded gray area); the remaining fibers ($\sim10\%$) are effectively super-virial ($m/m_\mathrm{vir} > 2$). Higher temperatures lower the line mass ratio by increasing the thermal component of $m_{\mathrm{vir}}$, with OMC-1 being the clearest case with a median value of $\sim0.3$. The number of structures with $m/m_\mathrm{vir}\gtrsim1$ per region appears to depend on the total mass of dense gas harboured in it, with OMC-2 and OMC-3 hosting the highest number of these virial fibers (see Table~\ref{tab:gen_prop}).

Although a minority, virial fibers seem to have a strong influence on the current star-formation properties of the regions in our sample. We can explore this connection through the number of protostars (P; see Sect.~\ref{subsec:n2h+}) harboured in the region and the protostars-over-disks ratio (P/D; see also Table~\ref{tab:sample_prop}). Figure~\ref{fig:gen_prop_plot} highlights how regions with continuous star formation and a significant number of protostars, such as OMC-2 and OMC-3 (P/D~$\gtrsim0.3$ and P~$\gtrsim25$; see Table~\ref{tab:sample_prop} and Paper I) are also those hosting a larger fraction of fibers with $m/m_\mathrm{vir}\gtrsim1$. We identified the star-forming fibers in our sample as those structures with protostars enclosed within less than one $\theta_\mathrm{beam}$ from the location of their constituent N$_2$H$^+$ components. According to this definition, one third of the fibers in the survey (50) is star forming with two thirds of the protostars in our fields (82) associated to them (our ALMA maps include 121 protostars, the majority of those covered by the IRAM-30m observations; see Paper I and Table~\ref{tab:sample_prop}).
We highlight these star-forming fibers in the M-L diagram (Figure~\ref{fig:mlpaper}, lower panel) by displaying them using colour-coded circles whose size is proportional to the number of protostars associated. For illustrative purposes, we also plot the expected value for $m_{\rm{vir}}$ at a constant temperature of $T_{\rm{K}}=20$~K (the typical gas temperature in our targets) and a non-thermal component $\sigma_\mathrm{nt}(L)$ following the standard scaling relation derived for filaments \citep[solid red line; see][]{hacar22}. Two qualitative results are drawn from this comparison: first, the star-forming fibers identified in our sample cluster towards the right side of the plot, approaching or even crossing the virial mass expected for filaments at 20~K. This behaviour suggests that the star-forming fibers are close to the virial condition, a hint further validated by the statistics of the individual regions. Figure~\ref{fig:virrat} shows the virial ratios for star-forming fibers (hatched histogram) and non-star-forming fibers (solid histogram) divided per region (colour-coded) and for the survey as a whole (grey histograms). The virial ratios for the star-forming fibers are consistently shifted towards higher values compared to those of non-star-forming fibers. This shift is reflected in the median values of $m/m_{\rm{vir}}$, which is $\langle m/m_{\rm{vir}}\rangle\sim1.1$ (dotted line) for the star-forming fibers and $\langle m/m_{\rm{vir}}\rangle\sim0.55$ (solid line) for the non-star-forming ones.
Second, and among the star-forming ones, fibers with higher line masses $m$ (see grey diagonal lines) usually show a higher number of protostars per object. We confirmed this trend again by computing the median virial ratio, weighted this time by the number of protostars. The resulting $\langle m/m_{\rm{vir}}\rangle\sim1.50$ (dashed line) is higher than the previous estimates suggesting a positive correlation between the line mass of the fiber and the number of protostars associated to it. In spite of some exceptions to this trend (e.g. OMC-1 South)\footnote{OMC-1 South is an outlier among the star-forming fibers given the number protostars associated to it (6) and its line mass ($m\sim37~\mathrm{M}_{\odot}~\mathrm{pc}^{-1}$). Because of these values, this region sits on the left side of the expected scaling relation for $m_\mathrm{vir}$ in Fig.~\ref{fig:mlpaper}, clearly departing from the suggested trend for star-forming fibers. OMC-1 South is an extremely dense and clustered region \citep[e.g.][]{palau18}, however, towards which N$_2$H$^+$ (7$-$6) shows emission as bright as, and even brighter than, N$_2$H$^+$ (1$-$0) \citep{hacar20sepia}. As a consequence, the column density and mass of H$_2$ inferred from N$_2$H$^+$ (1$-$0) may describe poorly the physical properties in the region. The departure of the region from the expected trend is therefore possibly more an observational bias than a physical differentiation.}, these star-forming fibers are the most massive structures in our EMERGE sample and are associated to known massive objects in our survey, such as OMC-2~FIR6 or OMC-3~MMS2. As a consequence, these fibers are found at the rightmost side of this distribution both in terms of $M$ and $m$. 

The ability of fibers to form stars was previously associated to their observed line mass. Fertile (or star-forming) fibers were identified as (super-)critical objects ($m/m_{\rm{crit}}\gtrsim 1$) in contrast to those sterile (or non star-forming) ones showing sub-critical line masses ($m/m_{\rm{crit}}\lesssim 1$) \citep[see][]{tafalla15,smith16}. As discussed before, however, the applicability of an idealised criterion based on the $m_{\rm{crit}}$ in a hydrostatic filament does not capture the complexity and dynamical nature of the filamentary structures revealed in the ISM at high spatial resolution. While still incomplete, the use of the virial ratio $m/m_{\rm{vir}}$ appears as better descriptor of the (global) potential of individual fibers to form young protostars according to Fig.~\ref{fig:mlpaper} (lower panel). The connection between the energy equipartition in a single fiber and the formation of individual cores and stars within will be the subject of future research in our project \citep{bonanomi25}.

\subsection{Fibers as the preferred organisation of the dense gas in clouds}\label{subsec:framework}

\begin{figure}[tbp]
    \centering
    \includegraphics[width=\linewidth]{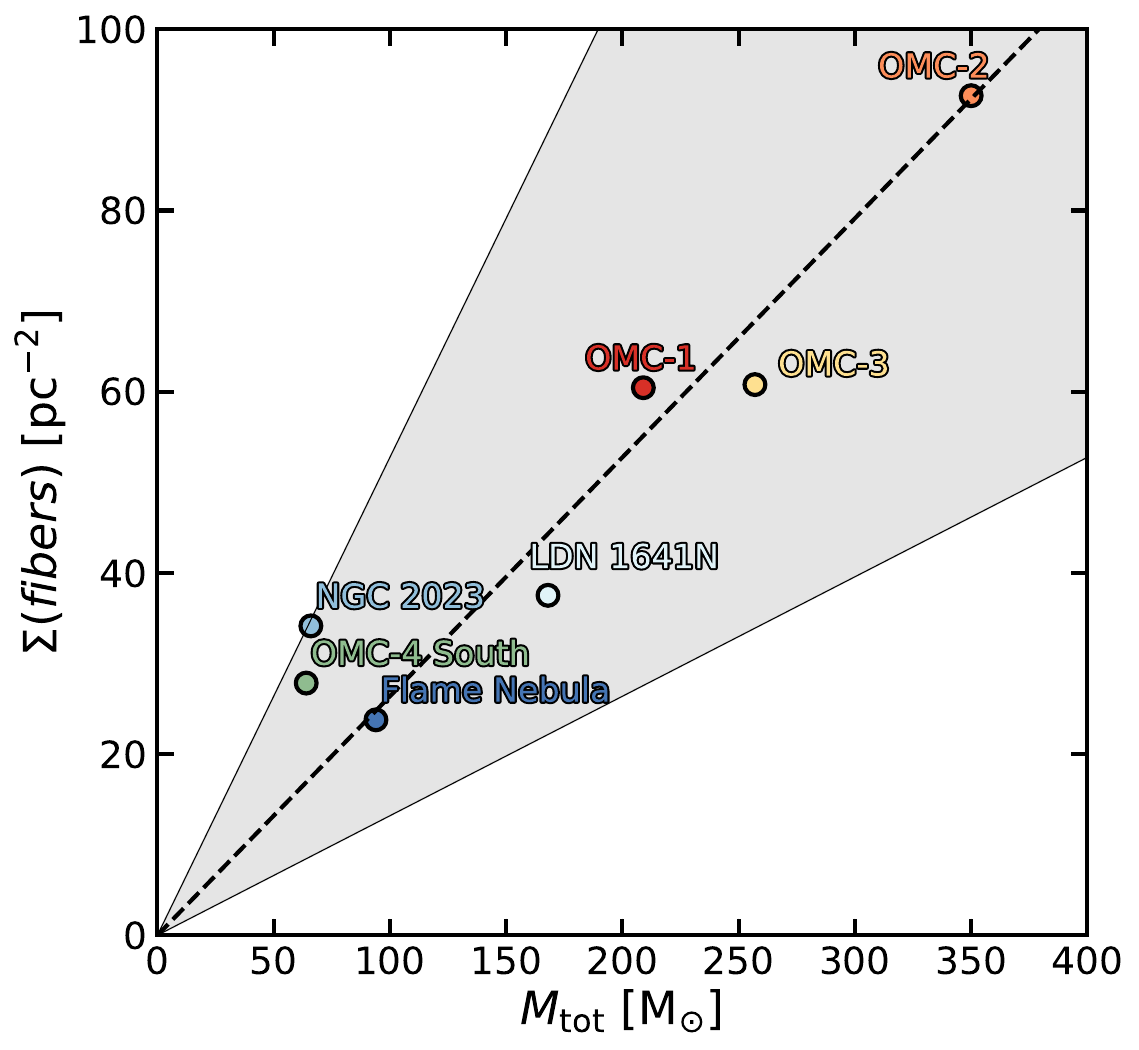}
    \caption{Comparison between the fiber surface density ($\Sigma(fibers)$) and total mass of dense gas ($M_\mathrm{tot}$) across the regions sampled in our EMERGE Early ALMA Survey. The black dashed line represents a linear fit of the data with $\Sigma \sim 0.26~M_\mathrm{tot}$. The gray shaded area is a factor of 2 confidence level for this correlation similar to the one estimated by \citet{hacar18}.}
    \label{fig:mlinvssurf}
\end{figure}

From the comparison of the fiber populations in five star-forming clouds, namely B213-L1495, Musca, NGC~1333, OMC-1 and OMC-2, \citet{hacar18} proposed a tentative linear correlation between the surface density of dense fibers ($\Sigma(fibers)$) and the (total) line mass ($m(tot)$) of these clouds. These results connect the internal gas structure across multiple star-formation environments suggesting fibers as the characteristic gas organisation within all type of low- (Musca and B213-L1495), intermediate- (NGC~1333 and OMC-2), and high-mass (OMC-1) star-forming regions. This smooth correlation implies that differences between these regions, as well as differences between isolated and clustered star-formation regimes, might be originated from the initial concentration of fibers in those same regions.
Although promising, the fiber populations investigated by Hacar et al. were characterised in star-forming regions at different distances (Taurus vs Orion) using different tracers (C$^{18}$O and N$_2$H$^+$) and resolutions (single-dish vs ALMA), limiting the conclusions of their work.

Our EMERGE ALMA Early survey allows us to further explore the above hypothesis, using this time 152 fibers homogeneously sampled and characterised with the same tracer and resolution by our observations in Orion.
Following Hacar et al., we calculated $\Sigma(fibers)$ diving the number of velocity-coherent structures identified by HiFIVe (Sect.~\ref{sec:fibersurvey}) by the area of our ALMA maps (Sect.~\ref{sec:observations}) for each region. Some of our regions appear more scattered and diffuse than elongated, however (e.g. OMC-4 South, NGC~2023; see Sect.~\ref{sec:densegas}). We therefore used the total dense gas mass in the region ($M_\mathrm{tot}$; see Table~\ref{tab:gen_prop}) instead of their total line mass ($m(dense)$) along the major axis of the map.
Figure~\ref{fig:mlinvssurf} displays the correlation between $\Sigma(fibers)$ and $M_\mathrm{tot}$ determined in our sample. The estimated $\Sigma(fibers)$ values again follow a linear correlation with $M_\mathrm{tot}$ throughout roughly one order of magnitude in both parameters. This correlation reads as $\Sigma (fibers) \sim 0.26~M_\mathrm{tot}$ according to a linear fit to the data. All regions in our survey reproduce this tight correlation within less than a factor of 2 difference (grey area) showing no discontinuity despite the different (proto-)stellar contents (OMC-1 vs OMC-4 South) or evolutionary stages (Flame Nebula vs NGC~2023). 
Instead, the above correlation indicates a direct proportionality between the current amount of dense gas in each region and the density of fibers detected in them. Or equivalently, fibers are generated in constant proportion per unit of dense gas available at the time.

As demonstrated in previous sections, the dense gas detected in our ALMA+IRAM-30m maps shows a systematic organisation ($>95\%$ of the spectra) into complex fiber networks in all the regions explored in our EMERGE Early ALMA Survey (Sect.~\ref{sec:observations}). Fibers in different regions share similar mass and length distributions (Sect.~\ref{subsec:MandL}). In all cases, these fibers exhibit a sonic, internal velocity dispersion ($\sigma_\mathrm{nt}/c_\mathrm{s}\sim 1$) irrespective of their star formation regime (low- and high-mass) or evolutionary stage (young and evolved) (Sect.~\ref{sec:observations}). The widespread presence of these objects in all type of regions suggest fibers as the preferred organisation of the dense gas in molecular clouds. The subsequent star formation activity of these clouds (e.g. number and density of protostars) naturally emerge from the global properties of these fiber systems (see above). In particular, clustered and high-mass star-forming regions are created in fiber arrangements of increasing surface density and complexity.

\section{Conclusions}\label{sec:conclusions}

We characterised the physical properties, kinematics, and virial condition of the EMERGE Early ALMA Survey, a sample of seven star-forming regions in Orion, comprising OMC-1, OMC-2, OMC-3, OMC-4 and LDN~1641N in Orion A, NGC~2023 and Flame Nebula in Orion B. Our sample includes different regions covering a wide range of star-formation regimes, cloud morphology, dense gas content, and evolutionary stages, all homogeneously surveyed in N$_2$H$^+$ (1$-$0) at 4.5$''$ (or 2000~au) using a series of large-scale ALMA+IRAM-30m mosaics (see also Paper I). Our main findings are listed below:

\begin{enumerate}
    \item Our N$_2$H$^+$ (1$-$0) observations sample the dense gas properties across three orders of magnitude in column density ($\sim2\times10^{21}-10^{24}$~cm$^{-2}$) throughout the survey. The dense gas traced by N$_2$H$^+$ is cold ($\sim 20$~K) in the majority of our fields and it is associated to most of the protostars in those same fields (Sects.~\ref{sec:densegas},~\ref{subsec:MandL}).
    
    \item Typically unresolved in previous studies, our high-resolution ALMA+IRAM-30m observations show a dense gas highly structured into densely populated networks of sub-parsec filamentary structures in all of our targets. Our results highlight the need for similar high-resolution observations to investigate the gas organisation prior to the formation of stars.
    
    \item Our analysis of the gas kinematics, traced in N$_2$H$^+$, identified 152 velocity-coherent fiber-like structures across our survey (Sect.~\ref{sec:fibersurvey}). These fibers share similar mass and length distributions, as well as sonic-like properties from low- to high-mass star-forming regions in our sample. The statistical significance of our sample demonstrates the pivotal role of fibers during the evolution of the dense gas in all prototypes of star-forming region.
    
    \item The identified fibers appear to have sub-virial ($m/m_\mathrm{vir}<1$) line masses on average. Yet, star-forming fibers (i.e. associated to protostars in the fields) show virial ratios $m/m_\mathrm{vir}\gtrsim1$, which is a factor of $\sim2-3$ higher than those of non-star-forming fibers (i.e. with no associated protostars). These star-forming fibers, despite being a few, seem to have a deep impact on the star-formation properties of the host region.

    \item By sharing many of their physical properties, fibers are arranged in networks of different complexity (Sect.~\ref{sec:densegas}). They are therefore the preferred organisational unit of the dense gas across low- to high-mass regions. This nature is confirmed by the tight linear correlation between the total mass of dense gas in a region and the surface density of fibers identified within. This approach follows and extends the previously reported linear correlation between the surface density of fibers and the total mass of dense gas found in nearby clouds (Sect.~\ref{subsec:framework}). The (current) star-formation properties (i.e. number and density of protostars) of our regions is then explained in terms of density of fibers observed in these same regions. 
\end{enumerate}

\section*{Data availability}
The table with all the results is available online via Zenodo at the following link: \url{https://doi.org/10.5281/zenodo.13628881}.\\
The ALMA+IRAM-30m N$_2$H$^+$ (1$-$0) cubes are instead available as Data Release 2 (DR2) on the following website: \url{https://emerge.univie.ac.at/results/data/}.

\begin{acknowledgements}
    The authors thank the referee for the thorough and extensive comments that helped improving considerably the manuscript.
    This project has received funding from the European Research Council (ERC) under the European Union’s Horizon 2020 research and innovation programme (Grant agreement No. 851435).
    M.T. acknowledges partial support from project PID2019-108765GB-I00 funded by MCIN/AEI/10.13039/501100011033.
    This paper makes use of the following ALMA data: ADS/JAO.ALMA\#2019.1.00641.S., ADS/JAO.ALMA\#2013.1.00662.S. ALMA is a partnership of ESO (representing its member states), NSF (USA) and NINS (Japan), together with NRC (Canada), MOST and ASIAA (Taiwan), and KASI (Republic of Korea), in cooperation with the Republic of Chile. The Joint ALMA Observatory is operated by ESO, AUI/NRAO and NAOJ.
    This work is based on IRAM-30m telescope observations carried out under project numbers 032-13, 120-20, 060-22, and 133-22. IRAM is supported by INSU/CNRS (France), MPG (Germany), and IGN (Spain). 
    This research has made use of the SIMBAD database, operated at CDS, Strasbourg, France.
    This research has made use of NASA’s Astrophysics Data System.
\end{acknowledgements}

\nocite{*}
\bibliographystyle{aa}
\bibliography{bibl.bib}

\begin{thebibliography}{84}
\expandafter\ifx\csname natexlab\endcsname\relax\def\natexlab#1{#1}\fi

\bibitem[{{Aikawa} {et~al.}(2005){Aikawa}, {Herbst}, {Roberts}, \& {Caselli}}]{aikawa05}
{Aikawa}, Y., {Herbst}, E., {Roberts}, H., \& {Caselli}, P. 2005, \apj, 620, 330

\bibitem[{{Andr{\'e}} {et~al.}(2014){Andr{\'e}}, {Di Francesco}, {Ward-Thompson}, {Inutsuka}, {Pudritz}, \& {Pineda}}]{andre14}
{Andr{\'e}}, P., {Di Francesco}, J., {Ward-Thompson}, D., {et~al.} 2014, in Protostars and Planets VI, ed. H.~{Beuther}, R.~S. {Klessen}, C.~P. {Dullemond}, \& T.~{Henning}, 27--51

\bibitem[{{Arzoumanian} {et~al.}(2019){Arzoumanian}, {Andr{\'e}}, {K{\"o}nyves}, {Palmeirim}, {Roy}, {Schneider}, {Benedettini}, {Didelon}, {Di Francesco}, {Kirk}, \& {Ladjelate}}]{arzoumanian19}
{Arzoumanian}, D., {Andr{\'e}}, P., {K{\"o}nyves}, V., {et~al.} 2019, \aap, 621, A42

\bibitem[{{Asplund} {et~al.}(2021){Asplund}, {Amarsi}, \& {Grevesse}}]{aspl21}
{Asplund}, M., {Amarsi}, A.~M., \& {Grevesse}, N. 2021, \aap, 653, A141

\bibitem[{{Bally}(2008)}]{bally08}
{Bally}, J. 2008, in Handbook of Star Forming Regions, Volume I, ed. B.~{Reipurth}, Vol.~4, 459

\bibitem[{{Bally} {et~al.}(1987){Bally}, {Langer}, {Stark}, \& {Wilson}}]{bally87}
{Bally}, J., {Langer}, W.~D., {Stark}, A.~A., \& {Wilson}, R.~W. 1987, \apjl, 312, L45

\bibitem[{{Barnard}(1907)}]{barnard07}
{Barnard}, E.~E. 1907, \apj, 25, 218

\bibitem[{{Barnes} {et~al.}(2021){Barnes}, {Henshaw}, {Fontani}, {Pineda}, {Cosentino}, {Tan}, {Caselli}, {Jim{\'e}nez-Serra}, {Law}, {Avison}, {Bigiel}, {Feng}, {Kong}, {Longmore}, {Moser}, {Parker}, {S{\'a}nchez-Monge}, \& {Wang}}]{barnes21}
{Barnes}, A.~T., {Henshaw}, J.~D., {Fontani}, F., {et~al.} 2021, \mnras, 503, 4601

\bibitem[{{Bergin} \& {Tafalla}(2007)}]{bergin07}
{Bergin}, E.~A. \& {Tafalla}, M. 2007, \araa, 45, 339

\bibitem[{{Bertoldi} \& {McKee}(1992)}]{bertoldi92}
{Bertoldi}, F. \& {McKee}, C.~F. 1992, \apj, 395, 140

\bibitem[{{Bohlin} {et~al.}(1978){Bohlin}, {Savage}, \& {Drake}}]{bohlin78}
{Bohlin}, R.~C., {Savage}, B.~D., \& {Drake}, J.~F. 1978, \apj, 224, 132

\bibitem[{{Bonanomi} {et~al.}(2025){Bonanomi}, {Hacar}, \& {Socci}}]{bonanomi25}
{Bonanomi}, F., {Hacar}, A., \& {Socci}, A. 2025, \aap

\bibitem[{{Bonanomi} {et~al.}(2024){Bonanomi}, {Hacar}, {Socci}, {Petry}, \& {Suri}}]{bonanomi24}
{Bonanomi}, F., {Hacar}, A., {Socci}, A., {Petry}, D., \& {Suri}, S. 2024, \aap, 688, A30

\bibitem[{{Caselli} {et~al.}(2002){Caselli}, {Benson}, {Myers}, \& {Tafalla}}]{caselli02}
{Caselli}, P., {Benson}, P.~J., {Myers}, P.~C., \& {Tafalla}, M. 2002, \apj, 572, 238

\bibitem[{{Caselli} {et~al.}(1995){Caselli}, {Myers}, \& {Thaddeus}}]{caselli95}
{Caselli}, P., {Myers}, P.~C., \& {Thaddeus}, P. 1995, \apjl, 455, L77

\bibitem[{{Chen} {et~al.}(2019){Chen}, {Zhang}, {Wright}, {Busquet}, {Lin}, {Liu}, {Olguin}, {Sanhueza}, {Nakamura}, {Palau}, {Ohashi}, {Tatematsu}, \& {Liao}}]{chen19}
{Chen}, H.-R.~V., {Zhang}, Q., {Wright}, M.~C.~H., {et~al.} 2019, \apj, 875, 24

\bibitem[{{Clarke} {et~al.}(2018){Clarke}, {Whitworth}, {Spowage}, {Duarte-Cabral}, {Suri}, {Jaffa}, {Walch}, \& {Clark}}]{clarke18}
{Clarke}, S.~D., {Whitworth}, A.~P., {Spowage}, R.~L., {et~al.} 2018, \mnras, 479, 1722

\bibitem[{{Dhabal} {et~al.}(2019){Dhabal}, {Mundy}, {Chen}, {Teuben}, \& {Storm}}]{dhab19}
{Dhabal}, A., {Mundy}, L.~G., {Chen}, C.-y., {Teuben}, P., \& {Storm}, S. 2019, \apj, 876, 108

\bibitem[{{Evans} {et~al.}(2009){Evans}, {Dunham}, {J{\o}rgensen}, {Enoch}, {Mer{\'\i}n}, {van Dishoeck}, {Alcal{\'a}}, {Myers}, {Stapelfeldt}, {Huard}, {Allen}, {Harvey}, {van Kempen}, {Blake}, {Koerner}, {Mundy}, {Padgett}, \& {Sargent}}]{evans09}
{Evans}, Neal~J., I., {Dunham}, M.~M., {J{\o}rgensen}, J.~K., {et~al.} 2009, \apjs, 181, 321

\bibitem[{{Furlan} {et~al.}(2016){Furlan}, {Fischer}, {Ali}, {Stutz}, {Stanke}, {Tobin}, {Megeath}, {Osorio}, {Hartmann}, {Calvet}, {Poteet}, {Booker}, {Manoj}, {Watson}, \& {Allen}}]{furlan16}
{Furlan}, E., {Fischer}, W.~J., {Ali}, B., {et~al.} 2016, \apjs, 224, 5

\bibitem[{{Gildas Team}(2013)}]{gil13}
{Gildas Team}. 2013, {GILDAS: Grenoble Image and Line Data Analysis Software}, Astrophysics Source Code Library, record ascl:1305.010

\bibitem[{{Goodman} {et~al.}(2014){Goodman}, {Alves}, {Beaumont}, {Benjamin}, {Borkin}, {Burkert}, {Dame}, {Jackson}, {Kauffmann}, {Robitaille}, \& {Smith}}]{goodman14}
{Goodman}, A.~A., {Alves}, J., {Beaumont}, C.~N., {et~al.} 2014, \apj, 797, 53

\bibitem[{{Goodman} {et~al.}(1998){Goodman}, {Barranco}, {Wilner}, \& {Heyer}}]{goodman98}
{Goodman}, A.~A., {Barranco}, J.~A., {Wilner}, D.~J., \& {Heyer}, M.~H. 1998, \apj, 504, 223

\bibitem[{{Hacar} {et~al.}(2020{\natexlab{a}}){Hacar}, {Bosman}, \& {van Dishoeck}}]{hacar20}
{Hacar}, A., {Bosman}, A.~D., \& {van Dishoeck}, E.~F. 2020{\natexlab{a}}, \aap, 635, A4

\bibitem[{{Hacar} {et~al.}(2023){Hacar}, {Clark}, {Heitsch}, {Kainulainen}, {Panopoulou}, {Seifried}, \& {Smith}}]{hacar22}
{Hacar}, A., {Clark}, S.~E., {Heitsch}, F., {et~al.} 2023, in Astronomical Society of the Pacific Conference Series, Vol. 534, Protostars and Planets VII, ed. S.~{Inutsuka}, Y.~{Aikawa}, T.~{Muto}, K.~{Tomida}, \& M.~{Tamura}, 153

\bibitem[{{Hacar} {et~al.}(2020{\natexlab{b}}){Hacar}, {Hogerheijde}, {Harsono}, {Portegies Zwart}, {De Breuck}, {Torstensson}, {Boland}, {Baryshev}, {Hesper}, {Barkhof}, {Adema}, {Bekema}, {Koops}, {Khudchenko}, \& {Stark}}]{hacar20sepia}
{Hacar}, A., {Hogerheijde}, M.~R., {Harsono}, D., {et~al.} 2020{\natexlab{b}}, \aap, 644, A133

\bibitem[{{Hacar} {et~al.}(2024){Hacar}, {Socci}, {Bonanomi}, {Petry}, {Tafalla}, {Harsono}, {Forbrich}, {Alves}, {Grossschedl}, {Goicoechea}, {Pety}, {Burkert}, \& {Li}}]{hacar24}
{Hacar}, A., {Socci}, A., {Bonanomi}, F., {et~al.} 2024, \aap, 687, A140

\bibitem[{{Hacar} \& {Tafalla}(2011)}]{hacar11}
{Hacar}, A. \& {Tafalla}, M. 2011, \aap, 533, A34

\bibitem[{{Hacar} {et~al.}(2017){Hacar}, {Tafalla}, \& {Alves}}]{hacar17}
{Hacar}, A., {Tafalla}, M., \& {Alves}, J. 2017, \aap, 606, A123

\bibitem[{{Hacar} {et~al.}(2018){Hacar}, {Tafalla}, {Forbrich}, {Alves}, {Meingast}, {Grossschedl}, \& {Teixeira}}]{hacar18}
{Hacar}, A., {Tafalla}, M., {Forbrich}, J., {et~al.} 2018, \aap, 610, A77

\bibitem[{{Hacar} {et~al.}(2013){Hacar}, {Tafalla}, {Kauffmann}, \& {Kov{\'a}cs}}]{hacar13}
{Hacar}, A., {Tafalla}, M., {Kauffmann}, J., \& {Kov{\'a}cs}, A. 2013, \aap, 554, A55

\bibitem[{{Hartmann}(2002)}]{hartmann02}
{Hartmann}, L. 2002, \apj, 578, 914

\bibitem[{{Henshaw} {et~al.}(2014){Henshaw}, {Caselli}, {Fontani}, {Jim{\'e}nez-Serra}, \& {Tan}}]{henshaw14}
{Henshaw}, J.~D., {Caselli}, P., {Fontani}, F., {Jim{\'e}nez-Serra}, I., \& {Tan}, J.~C. 2014, \mnras, 440, 2860

\bibitem[{{Herbst} {et~al.}(2000){Herbst}, {Terzieva}, \& {Talbi}}]{herbst00}
{Herbst}, E., {Terzieva}, R., \& {Talbi}, D. 2000, \mnras, 311, 869

\bibitem[{{Howard} {et~al.}(2019){Howard}, {Whitworth}, {Marsh}, {Clarke}, {Griffin}, {Smith}, \& {Lomax}}]{howard19}
{Howard}, A.~D.~P., {Whitworth}, A.~P., {Marsh}, K.~A., {et~al.} 2019, \mnras, 489, 962

\bibitem[{{Huchra} \& {Geller}(1982)}]{huchra82}
{Huchra}, J.~P. \& {Geller}, M.~J. 1982, \apj, 257, 423

\bibitem[{{Inutsuka} \& {Miyama}(1997)}]{inut97}
{Inutsuka}, S.-i. \& {Miyama}, S.~M. 1997, \apj, 480, 681

\bibitem[{{Jackson} {et~al.}(2010){Jackson}, {Finn}, {Chambers}, {Rathborne}, \& {Simon}}]{jackson10}
{Jackson}, J.~M., {Finn}, S.~C., {Chambers}, E.~T., {Rathborne}, J.~M., \& {Simon}, R. 2010, \apjl, 719, L185

\bibitem[{{Johnstone} \& {Bally}(1999)}]{john99}
{Johnstone}, D. \& {Bally}, J. 1999, \apjl, 510, L49

\bibitem[{{Johnstone} \& {Bally}(2006)}]{johnstone06}
{Johnstone}, D. \& {Bally}, J. 2006, \apj, 653, 383

\bibitem[{{Kainulainen} {et~al.}(2017){Kainulainen}, {Stutz}, {Stanke}, {Abreu-Vicente}, {Beuther}, {Henning}, {Johnston}, \& {Megeath}}]{kainul17}
{Kainulainen}, J., {Stutz}, A.~M., {Stanke}, T., {et~al.} 2017, \aap, 600, A141

\bibitem[{{Kauffmann} {et~al.}(2008){Kauffmann}, {Bertoldi}, {Bourke}, {Evans}, \& {Lee}}]{kauff08}
{Kauffmann}, J., {Bertoldi}, F., {Bourke}, T.~L., {Evans}, N.~J., I., \& {Lee}, C.~W. 2008, \aap, 487, 993

\bibitem[{{Klessen} {et~al.}(2000){Klessen}, {Heitsch}, \& {Mac Low}}]{klessen00}
{Klessen}, R.~S., {Heitsch}, F., \& {Mac Low}, M.-M. 2000, \apj, 535, 887

\bibitem[{{Kong} {et~al.}(2018){Kong}, {Arce}, {Feddersen}, {Carpenter}, {Nakamura}, {Shimajiri}, {Isella}, {Ossenkopf-Okada}, {Sargent}, {S{\'a}nchez-Monge}, {Suri}, {Kauffmann}, {Pillai}, {Pineda}, {Koda}, {Bally}, {Lis}, {Padoan}, {Klessen}, {Mairs}, {Goodman}, {Goldsmith}, {McGehee}, {Schilke}, {Teuben}, {Maureira}, {Hara}, {Ginsburg}, {Burkhart}, {Smith}, {Schmiedeke}, {Pineda}, {Ishii}, {Sasaki}, {Kawabe}, {Urasawa}, {Oyamada}, \& {Tanabe}}]{Kong2018}
{Kong}, S., {Arce}, H.~G., {Feddersen}, J.~R., {et~al.} 2018, \apjs, 236, 25

\bibitem[{{K{\"o}nyves} {et~al.}(2020){K{\"o}nyves}, {Andr{\'e}}, {Arzoumanian}, {Schneider}, {Men'shchikov}, {Bontemps}, {Ladjelate}, {Didelon}, {Pezzuto}, {Benedettini}, {Bracco}, {Di Francesco}, {Goodwin}, {Rygl}, {Shimajiri}, {Spinoglio}, {Ward-Thompson}, \& {White}}]{konyves20}
{K{\"o}nyves}, V., {Andr{\'e}}, P., {Arzoumanian}, D., {et~al.} 2020, \aap, 635, A34

\bibitem[{{Krumholz} \& {McKee}(2008)}]{krum08}
{Krumholz}, M.~R. \& {McKee}, C.~F. 2008, \nat, 451, 1082

\bibitem[{{Lee} {et~al.}(2014){Lee}, {Fern{\'a}ndez-L{\'o}pez}, {Storm}, {Looney}, {Mundy}, {Segura-Cox}, {Teuben}, {Rosolowsky}, {Arce}, {Ostriker}, {Shirley}, {Kwon}, {Kauffmann}, {Tobin}, {Plunkett}, {Pound}, {Salter}, {Volgenau}, {Chen}, {Tassis}, {Isella}, {Crutcher}, {Gammie}, \& {Testi}}]{lee14}
{Lee}, K.~I., {Fern{\'a}ndez-L{\'o}pez}, M., {Storm}, S., {et~al.} 2014, \apj, 797, 76

\bibitem[{{Li} {et~al.}(2013){Li}, {Kauffmann}, {Zhang}, \& {Chen}}]{li13}
{Li}, D., {Kauffmann}, J., {Zhang}, Q., \& {Chen}, W. 2013, \apjl, 768, L5

\bibitem[{{Lombardi} {et~al.}(2014){Lombardi}, {Bouy}, {Alves}, \& {Lada}}]{lombardi14}
{Lombardi}, M., {Bouy}, H., {Alves}, J., \& {Lada}, C.~J. 2014, \aap, 566, A45

\bibitem[{{Lynds}(1962)}]{lynds62}
{Lynds}, B.~T. 1962, \apjs, 7, 1

\bibitem[{{Mairs} {et~al.}(2016){Mairs}, {Johnstone}, {Kirk}, {Buckle}, {Berry}, {Broekhoven-Fiene}, {Currie}, {Fich}, {Graves}, {Hatchell}, {Jenness}, {Mottram}, {Nutter}, {Pattle}, {Pineda}, {Salji}, {Di Francesco}, {Hogerheijde}, {Ward-Thompson}, {Bastien}, {Bresnahan}, {Butner}, {Chen}, {Chrysostomou}, {Coud{\'e}}, {Davis}, {Drabek-Maunder}, {Duarte-Cabral}, {Fiege}, {Friberg}, {Friesen}, {Fuller}, {Greaves}, {Gregson}, {Holland}, {Joncas}, {Kirk}, {Knee}, {Marsh}, {Matthews}, {Moriarty-Schieven}, {Mowat}, {Rawlings}, {Richer}, {Robertson}, {Rosolowsky}, {Rumble}, {Sadavoy}, {Thomas}, {Tothill}, {Viti}, {White}, {Wouterloot}, {Yates}, \& {Zhu}}]{mairs16}
{Mairs}, S., {Johnstone}, D., {Kirk}, H., {et~al.} 2016, \mnras, 461, 4022

\bibitem[{{Megeath} {et~al.}(2012){Megeath}, {Gutermuth}, {Muzerolle}, {Kryukova}, {Flaherty}, {Hora}, {Allen}, {Hartmann}, {Myers}, {Pipher}, {Stauffer}, {Young}, \& {Fazio}}]{megeath12}
{Megeath}, S.~T., {Gutermuth}, R., {Muzerolle}, J., {et~al.} 2012, \aj, 144, 192

\bibitem[{{Megeath} {et~al.}(2016){Megeath}, {Gutermuth}, {Muzerolle}, {Kryukova}, {Hora}, {Allen}, {Flaherty}, {Hartmann}, {Myers}, {Pipher}, {Stauffer}, {Young}, \& {Fazio}}]{megeath16}
{Megeath}, S.~T., {Gutermuth}, R., {Muzerolle}, J., {et~al.} 2016, \aj, 151, 5

\bibitem[{{Menten} {et~al.}(2007){Menten}, {Reid}, {Forbrich}, \& {Brunthaler}}]{menten07}
{Menten}, K.~M., {Reid}, M.~J., {Forbrich}, J., \& {Brunthaler}, A. 2007, \aap, 474, 515

\bibitem[{{Miesch} \& {Bally}(1994)}]{miesch94}
{Miesch}, M.~S. \& {Bally}, J. 1994, \apj, 429, 645

\bibitem[{{Molinari} {et~al.}(2010){Molinari}, {Swinyard}, {Bally}, {Barlow}, {Bernard}, {Martin}, {Moore}, {Noriega-Crespo}, {Plume}, {Testi}, {Zavagno}, {Abergel}, {Ali}, {Anderson}, {Andr{\'e}}, {Baluteau}, {Battersby}, {Beltr{\'a}n}, {Benedettini}, {Billot}, {Blommaert}, {Bontemps}, {Boulanger}, {Brand}, {Brunt}, {Burton}, {Calzoletti}, {Carey}, {Caselli}, {Cesaroni}, {Cernicharo}, {Chakrabarti}, {Chrysostomou}, {Cohen}, {Compiegne}, {de Bernardis}, {de Gasperis}, {di Giorgio}, {Elia}, {Faustini}, {Flagey}, {Fukui}, {Fuller}, {Ganga}, {Garcia-Lario}, {Glenn}, {Goldsmith}, {Griffin}, {Hoare}, {Huang}, {Ikhenaode}, {Joblin}, {Joncas}, {Juvela}, {Kirk}, {Lagache}, {Li}, {Lim}, {Lord}, {Marengo}, {Marshall}, {Masi}, {Massi}, {Matsuura}, {Minier}, {Miville-Desch{\^e}nes}, {Montier}, {Morgan}, {Motte}, {Mottram}, {M{\"u}ller}, {Natoli}, {Neves}, {Olmi}, {Paladini}, {Paradis}, {Parsons}, {Peretto}, {Pestalozzi}, {Pezzuto}, {Piacentini}, {Piazzo}, {Polychroni}, {Pomar{\`e}s}, {Popescu}, {Reach}, {Ristorcelli},
  {Robitaille}, {Robitaille}, {Rod{\'o}n}, {Roy}, {Royer}, {Russeil}, {Saraceno}, {Sauvage}, {Schilke}, {Schisano}, {Schneider}, {Schuller}, {Schulz}, {Sibthorpe}, {Smith}, {Smith}, {Spinoglio}, {Stamatellos}, {Strafella}, {Stringfellow}, {Sturm}, {Taylor}, {Thompson}, {Traficante}, {Tuffs}, {Umana}, {Valenziano}, {Vavrek}, {Veneziani}, {Viti}, {Waelkens}, {Ward-Thompson}, {White}, {Wilcock}, {Wyrowski}, {Yorke}, \& {Zhang}}]{molinari10}
{Molinari}, S., {Swinyard}, B., {Bally}, J., {et~al.} 2010, \aap, 518, L100

\bibitem[{{Myers}(2009)}]{myers09}
{Myers}, P.~C. 2009, \apj, 700, 1609

\bibitem[{{Nishimura} {et~al.}(2015){Nishimura}, {Tokuda}, {Kimura}, {Muraoka}, {Maezawa}, {Ogawa}, {Dobashi}, {Shimoikura}, {Mizuno}, {Fukui}, \& {Onishi}}]{Nishimura2015}
{Nishimura}, A., {Tokuda}, K., {Kimura}, K., {et~al.} 2015, \apjs, 216, 18

\bibitem[{{Orkisz} {et~al.}(2019){Orkisz}, {Peretto}, {Pety}, {Gerin}, {Levrier}, {Bron}, {Bardeau}, {Goicoechea}, {Gratier}, {Guzm{\'a}n}, {Hughes}, {Languignon}, {Le Petit}, {Liszt}, {{\"O}berg}, {Roueff}, {Sievers}, \& {Tremblin}}]{ork19}
{Orkisz}, J.~H., {Peretto}, N., {Pety}, J., {et~al.} 2019, \aap, 624, A113

\bibitem[{{Ostriker}(1964)}]{ostriker64}
{Ostriker}, J. 1964, \apj, 140, 1056

\bibitem[{{Palau} {et~al.}(2018){Palau}, {Zapata}, {Rom{\'a}n-Z{\'u}{\~n}iga}, {S{\'a}nchez-Monge}, {Estalella}, {Busquet}, {Girart}, {Fuente}, \& {Commer{\c{c}}on}}]{palau18}
{Palau}, A., {Zapata}, L.~A., {Rom{\'a}n-Z{\'u}{\~n}iga}, C.~G., {et~al.} 2018, \apj, 855, 24

\bibitem[{{Palmeirim} {et~al.}(2013){Palmeirim}, {Andr{\'e}}, {Kirk}, {Ward-Thompson}, {Arzoumanian}, {K{\"o}nyves}, {Didelon}, {Schneider}, {Benedettini}, {Bontemps}, {Di Francesco}, {Elia}, {Griffin}, {Hennemann}, {Hill}, {Martin}, {Men'shchikov}, {Molinari}, {Motte}, {Nguyen Luong}, {Nutter}, {Peretto}, {Pezzuto}, {Roy}, {Rygl}, {Spinoglio}, \& {White}}]{palme13}
{Palmeirim}, P., {Andr{\'e}}, P., {Kirk}, J., {et~al.} 2013, \aap, 550, A38

\bibitem[{{Peterson} \& {Megeath}(2008)}]{peterson08}
{Peterson}, D.~E. \& {Megeath}, S.~T. 2008, in Handbook of Star Forming Regions, Volume I, ed. B.~{Reipurth}, Vol.~4, 590

\bibitem[{{Pety}(2005)}]{pety05}
{Pety}, J. 2005, in SF2A-2005: Semaine de l'Astrophysique Francaise, ed. F.~{Casoli}, T.~{Contini}, J.~M. {Hameury}, \& L.~{Pagani}, 721

\bibitem[{{Priestley} {et~al.}(2023){Priestley}, {Arzoumanian}, \& {Whitworth}}]{Priestley2023}
{Priestley}, F.~D., {Arzoumanian}, D., \& {Whitworth}, A.~P. 2023, \mnras, 522, 3890

\bibitem[{{Rivilla} {et~al.}(2013){Rivilla}, {Mart{\'\i}n-Pintado}, {Sanz-Forcada}, {Jim{\'e}nez-Serra}, \& {Rodr{\'\i}guez-Franco}}]{rivilla13}
{Rivilla}, V.~M., {Mart{\'\i}n-Pintado}, J., {Sanz-Forcada}, J., {Jim{\'e}nez-Serra}, I., \& {Rodr{\'\i}guez-Franco}, A. 2013, \mnras, 434, 2313

\bibitem[{{Schneider} \& {Elmegreen}(1979)}]{schneider79}
{Schneider}, S. \& {Elmegreen}, B.~G. 1979, \apjs, 41, 87

\bibitem[{{Schuller} {et~al.}(2021){Schuller}, {Andr{\'e}}, {Shimajiri}, {Zavagno}, {Peretto}, {Arzoumanian}, {Csengeri}, {K{\"o}nyves}, {Palmeirim}, {Pezzuto}, {Rigby}, {Roussel}, {Ajeddig}, {Dumaye}, {Gallais}, {Le Pennec}, {Martignac}, {Mattern}, {Rev{\'e}ret}, {Rodriguez}, \& {Talvard}}]{schuller21}
{Schuller}, F., {Andr{\'e}}, P., {Shimajiri}, Y., {et~al.} 2021, \aap, 651, A36

\bibitem[{{Shimajiri} {et~al.}(2023){Shimajiri}, {Andr{\'e}}, {Peretto}, {Arzoumanian}, {Ntormousi}, \& {K{\"o}nyves}}]{shima23}
{Shimajiri}, Y., {Andr{\'e}}, P., {Peretto}, N., {et~al.} 2023, \aap, 672, A133

\bibitem[{{Shimajiri} {et~al.}(2014){Shimajiri}, {Kitamura}, {Saito}, {Momose}, {Nakamura}, {Dobashi}, {Shimoikura}, {Nishitani}, {Yamabi}, {Hara}, {Katakura}, {Tsukagoshi}, {Tanaka}, \& {Kawabe}}]{Shimajiri2014}
{Shimajiri}, Y., {Kitamura}, Y., {Saito}, M., {et~al.} 2014, \aap, 564, A68

\bibitem[{{Smith} {et~al.}(2016){Smith}, {Glover}, {Klessen}, \& {Fuller}}]{smith16}
{Smith}, R.~J., {Glover}, S. C.~O., {Klessen}, R.~S., \& {Fuller}, G.~A. 2016, \mnras, 455, 3640

\bibitem[{{Socci} {et~al.}(2024){Socci}, {Hacar}, {Bonanomi}, {Tafalla}, \& {Suri}}]{socci24}
{Socci}, A., {Hacar}, A., {Bonanomi}, F., {Tafalla}, M., \& {Suri}, S. 2024, \aap

\bibitem[{{Sokolov} {et~al.}(2019){Sokolov}, {Wang}, {Pineda}, {Caselli}, {Henshaw}, {Barnes}, {Tan}, {Fontani}, \& {Jim{\'e}nez-Serra}}]{sokolov19}
{Sokolov}, V., {Wang}, K., {Pineda}, J.~E., {et~al.} 2019, \apj, 872, 30

\bibitem[{{Stanke} {et~al.}(2022){Stanke}, {Arce}, {Bally}, {Bergman}, {Carpenter}, {Davis}, {Dent}, {Di Francesco}, {Eisl{\"o}ffel}, {Froebrich}, {Ginsburg}, {Heyer}, {Johnstone}, {Mardones}, {McCaughrean}, {Megeath}, {Nakamura}, {Smith}, {Stutz}, {Tatematsu}, {Walker}, {Williams}, {Zinnecker}, {Swift}, {Kulesa}, {Peters}, {Duffy}, {Kloosterman}, {Y{\ensuremath{\i}}ld{\ensuremath{\i}}z}, {Pineda}, {De Breuck}, \& {Klein}}]{stanke22}
{Stanke}, T., {Arce}, H.~G., {Bally}, J., {et~al.} 2022, \aap, 658, A178

\bibitem[{{Stutz} {et~al.}(2013){Stutz}, {Tobin}, {Stanke}, {Megeath}, {Fischer}, {Robitaille}, {Henning}, {Ali}, {di Francesco}, {Furlan}, {Hartmann}, {Osorio}, {Wilson}, {Allen}, {Krause}, \& {Manoj}}]{stutz13}
{Stutz}, A.~M., {Tobin}, J.~J., {Stanke}, T., {et~al.} 2013, \apj, 767, 36

\bibitem[{{Suri} {et~al.}(2019){Suri}, {S{\'a}nchez-Monge}, {Schilke}, {Clarke}, {Smith}, {Ossenkopf-Okada}, {Klessen}, {Padoan}, {Goldsmith}, {Arce}, {Bally}, {Carpenter}, {Ginsburg}, {Johnstone}, {Kauffmann}, {Kong}, {Lis}, {Mairs}, {Pillai}, {Pineda}, \& {Duarte-Cabral}}]{suri19}
{Suri}, S., {S{\'a}nchez-Monge}, {\'A}., {Schilke}, P., {et~al.} 2019, \aap, 623, A142

\bibitem[{{Tafalla} \& {Hacar}(2015)}]{tafalla15}
{Tafalla}, M. \& {Hacar}, A. 2015, \aap, 574, A104

\bibitem[{{Tafalla} {et~al.}(2023){Tafalla}, {Usero}, \& {Hacar}}]{tafalla23}
{Tafalla}, M., {Usero}, A., \& {Hacar}, A. 2023, \aap, 679, A112

\bibitem[{{Tan} {et~al.}(2014){Tan}, {Beltr{\'a}n}, {Caselli}, {Fontani}, {Fuente}, {Krumholz}, {McKee}, \& {Stolte}}]{tan14}
{Tan}, J.~C., {Beltr{\'a}n}, M.~T., {Caselli}, P., {et~al.} 2014, in Protostars and Planets VI, ed. H.~{Beuther}, R.~S. {Klessen}, C.~P. {Dullemond}, \& T.~{Henning}, 149--172

\bibitem[{{Teixeira} {et~al.}(2016){Teixeira}, {Takahashi}, {Zapata}, \& {Ho}}]{teixeira16}
{Teixeira}, P.~S., {Takahashi}, S., {Zapata}, L.~A., \& {Ho}, P.~T.~P. 2016, \aap, 587, A47

\bibitem[{{Wenger} {et~al.}(2000){Wenger}, {Ochsenbein}, {Egret}, {Dubois}, {Bonnarel}, {Borde}, {Genova}, {Jasniewicz}, {Lalo{\"e}}, {Lesteven}, \& {Monier}}]{wenger00}
{Wenger}, M., {Ochsenbein}, F., {Egret}, D., {et~al.} 2000, \aaps, 143, 9

\bibitem[{{Wiseman} \& {Ho}(1998)}]{wise98}
{Wiseman}, J.~J. \& {Ho}, P. T.~P. 1998, \apj, 502, 676

\bibitem[{{Yun} {et~al.}(2021){Yun}, {Lee}, {Choi}, {Evans}, {Offner}, {Heyer}, {Gaches}, {Lee}, {Baek}, {Choi}, {Kang}, {Lee}, {Tatematsu}, {Yang}, {Chen}, {Lee}, {Jung}, {Lee}, \& {Cho}}]{yun21}
{Yun}, H.-S., {Lee}, J.-E., {Choi}, Y., {et~al.} 2021, \apjs, 256, 16

\bibitem[{{Zamora-Avil{\'e}s} {et~al.}(2017){Zamora-Avil{\'e}s}, {Ballesteros-Paredes}, \& {Hartmann}}]{zamora17}
{Zamora-Avil{\'e}s}, M., {Ballesteros-Paredes}, J., \& {Hartmann}, L.~W. 2017, \mnras, 472, 647

\end{thebibliography}



\begin{appendix}

\section{Additional maps and tables}\label{sec:mapsandtab}

\begin{figure*}[tbp]
  \centering
  \captionsetup{format=overlay, labelsep=frcolon}
  \caption{}
  \begin{tikzpicture}
    \node[anchor=south west,inner sep=0] (image) at (0,0)
    {\includegraphics[width=0.98\linewidth]{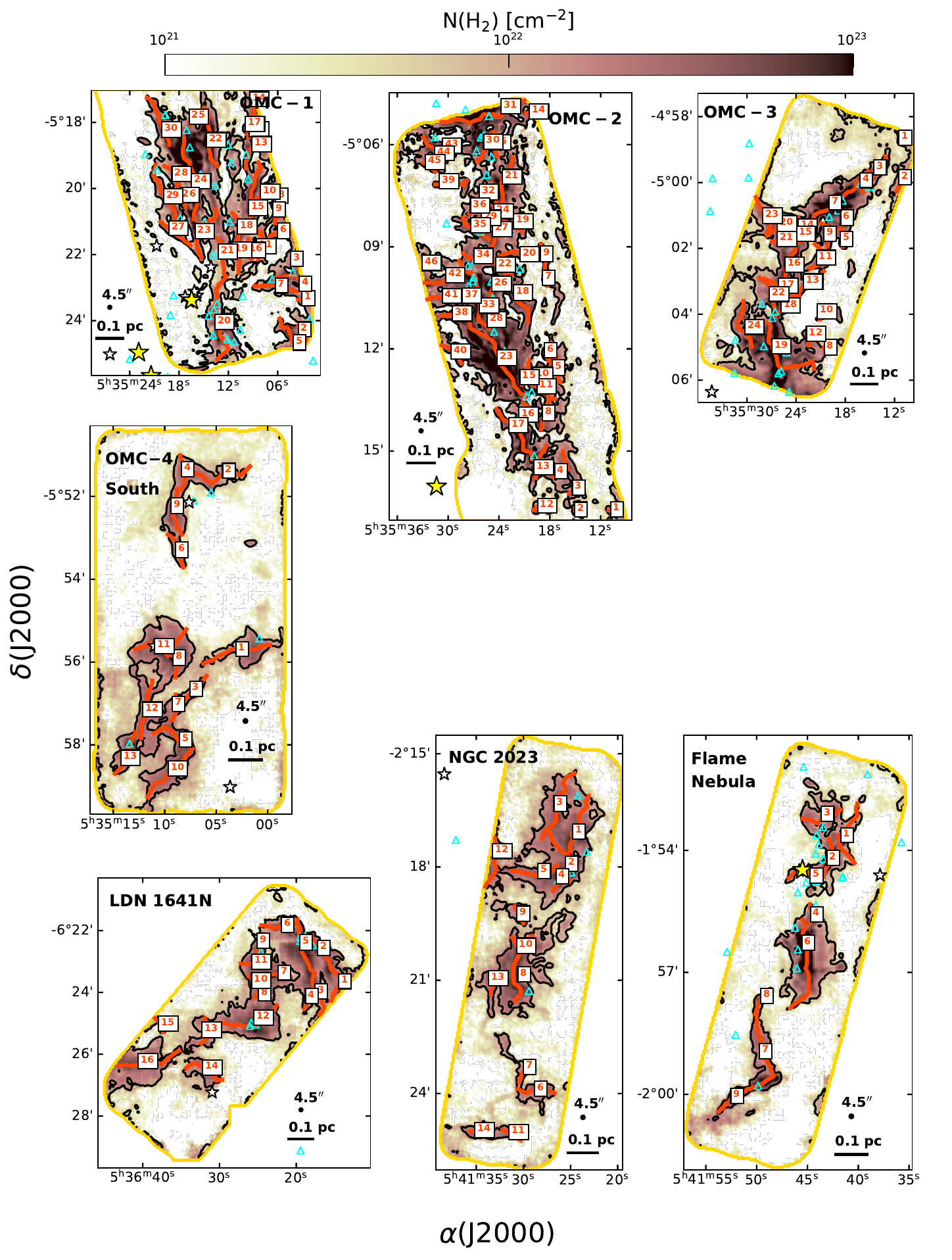}};
    \begin{scope}[x={(image.south east)},y={(image.north west)}]
      \draw let \p1 = (0.67,0.5)
      in node[text width=\x1,align=center,color=black] at
      (0.67, 0.5) {\textbf{Fig.~A.1:}  Column density maps of the 7 star-forming regions composing the EMERGE Early ALMA Survey. Same as Fig. 1-7, the cyan triangles are the protostars, the yellow and white stars the OB stars and the red segments the axes of our fibers. Each one of these fibers is associated here with its ID derived by HiFIVe.};
    \end{scope}
  \end{tikzpicture}
  \label{fig:numberedfib}
\end{figure*}

\begin{figure*}[tbp]
    \centering
    \includegraphics[width=\linewidth]{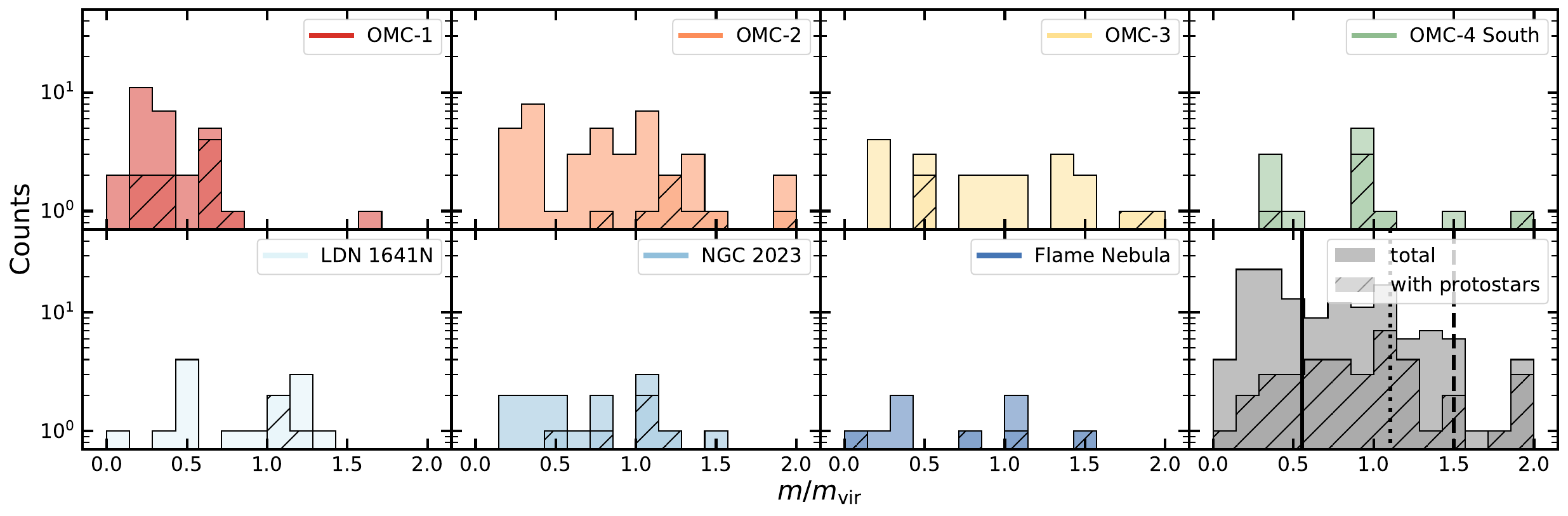}
    \caption{Virial ratios of the Orion fibers obtained per region in our survey (see Sect.~\ref{subsec:linemasses}). The fibers are further divided in star-forming (with protostars, hatched histogram) and total (with and without protostars, plain histogram). The grey histogram reports the virial ratios for the survey as a whole, along with the median values for the non-star-forming fibers (solid line) and the star-forming ones (dashed line).}
    \label{fig:virrat}
\end{figure*}

In this Appendix, we focus more on the individual fibers composing the survey, instead of their statistical properties as a whole. To this end, we provide the physical properties and the spatial distribution of each fiber among the 152 identified in the EMERGE Early ALMA Survey are found online via Zenodo. Each one of the fibers is labeled with an ID, assigned by HiFIVe based on the mean x-position in the map, therefore, from West to East in the maps. The corresponding fibers are displayed in Figure~\ref{fig:numberedfib} over the column density maps of the EMERGE Early ALMA Survey (see also Fig.~\ref{fig:NH2}) with each one bearing its own ID.

When looking at the single virial ratios throughout the survey, we note some features. All regions show different distributions of the virial ratios, and a corresponding variability in their further division between star-forming and non-star-forming fibers. Regions such as OMC-1 show a higher contribution from both star-forming and non-star-forming fibers for virial ratios $m/m_\mathrm{vir}\lesssim1$ on average in the survey. This feature is likely due to a high degree of fragmentation in the region (30 fibers) and the substantial contribution of the thermal component ($T_\mathrm{K}\gtrsim25~K$ on average) to $m_\mathrm{vir}$ (see Eq.~\ref{mvirial}). Regions such as OMC-2 show instead a clearer shift between the virial ratios of non-star-forming and star-forming fibers, with the ratios of the latter showing higher values on average. Despite the variety of distributions, a systematic shift of the star-forming fiber towards higher virial ratios compared to the non-star-forming ones, within the same region, is seen throughout the survey. This behaviour, confirmed by the different median values (see Fig.~\ref{fig:virrat} and Sect.~\ref{subsec:linemasses}), highlights the importance of higher line masses in connection to the star-formation activity within a single fiber\footnote{We note that, while the absolute value for the virial ratio may be debated between 1D and 3D (with a factor of 3 difference in Eq.~\ref{mvirial}), the shift in median virial ratio between star-forming and non-star-forming fibers is not affected by a scaling factor and bears a deeper meaning than a stability condition for the structure.}.

\section{Semi-automatic analysis of the EMERGE Early ALMA Survey}\label{sec:filchap}

In this Appendix, we took the points outlined in Sect.~\ref{sec:analysis} and expanded on the main algorithmic choices made during the analysis and its main outputs.

\subsection{Spectral fitting}\label{subsec:manualfit}

The fitting of the N$_2$H$^+$ (1$-$0) lines was carried out in groups of $5\times5$ spectra with supervised GILDAS/CLASS customised routines \citep{pety05,gil13}. The hyperfine structure of the line \citep[e.g.][]{caselli95} was fitted accordingly using the \texttt{hfs} method \citep[see GILDAS guidelines\footnote{\url{https://www.iram.fr/IRAMFR/GILDAS/doc/html/class-html/node11.html}};][]{pety05,gil13}, providing information on its antenna temperature ($T_\mathrm{ant}\times\tau$), total opacity ($\tau_\mathrm{main}$), linewidth ($\Delta v$), centroid velocity ($V_\mathrm{lsr}$), and noise level ($\sigma$). These parameters were further used to derive other spectral properties, such as the peak temperature ($T_\mathrm{peak}$) and the integrated intensity ($I$). Different selection criteria ensured the quality and signal-to-noise (S/N) of the spectra as well as the optimisation of the process \citep[see also Appendix B in][]{hacar18}. We also required our fits to fulfill the following criteria: (a) $2\le V_\mathrm{lsr} \le 15$, the velocity range in Orion \citep[e.g.][]{bally08}; (b) $I > 2.5~\rm{K~km~s}^{-1}$, a threshold which roughly corresponds to the floor term in Eq.~(\ref{eq:caln2hp}); (c) $T_\mathrm{peak}\geq3\sigma$, or $\mathrm{S/N}\geq3$ at the peak. The noise is determined directly from the fitting of a baseline in each spectral channel. Throughout our survey the noise is homogeneous with a typical value of $\sigma\sim0.020$~Jy~beam$^{-1}$, or $\sigma\sim0.15$~K after applying the Jy-to-K conversion (see below); (d) $\Delta v > \Delta V_\mathrm{th}$, the thermal broadening of the line computed as $\sqrt{8~\mathrm{log}(2)~k_\mathrm{B}~T_\mathrm{K}/\mu_\mathrm{N_2H^+}}$, with $k_\mathrm{B}$ Boltzmann constant and $\mu_\mathrm{N_2H^+}$ molecular mass of N$_2$H$^+$. In all cases we applied standard Jy-to-K conversions (e.g. from fluxes in Jy~beam~km~s$^{-1}$ into K~km~s$^{-1}$ in main beam temperature $T_\mathrm{mb}$ units; see Paper I) and conversions between line intensities and total gas column densities (N(H$_2$)) by applying Eq.~(\ref{eq:caln2hp}). As discussed in Sect.~\ref{sec:analysis}, the selection retrieved a total of $\sim57,000$ independent velocity components available for the analysis out of the initial of $\sim170,000$ N$_2$H$^+$ spectra sampled in our EMERGE Early ALMA survey.

\subsection{Identification of velocity-coherent structures}\label{subsec:identification}

A quantitative description of the complex dense gas morphology seen in Fig.~\ref{fig:NH2} required its decomposition in structures and the determination of their physical properties. In particular, parsec-scale filaments observed in our Galaxy \citep[e.g. B213/L1495;][]{palme13} are composed by associations of velocity-coherent sub-filaments \citep{hacar13}. The continuity in space and velocity \citep[i.e. velocity coherence;][]{goodman98} of these so-called fibers \citep{andre14} is a key property of these objects. They are in fact thought to be the final hierarchy among different filament families before the transition to cores, which would inherit their kinematic properties \citep[see][for a review]{hacar22}. 

To capture both the hierarchical nature of our filamentary clouds and their decomposition into velocity-coherent structures, we use the Hierarchical Friends in Velocity (HiFIVe) algorithm, previously employed in similar studies \citep{hacar18}. HiFIVe is devised to identify structures which are continuous in the position-position-velocity (PPV) plane using a Friends-of-Friends (FoF) approach \citep{huchra82}: each field is progressively associated to its nearest neighbours within a maximum separation (linking metric) in both space ($\theta_\mathrm{beam}$) and velocity ($\Delta v/2\times 2d/\theta_\mathrm{beam}$, where $d$ is the distance between fields in arcsec). This identification process and the growth of structures through the progressive association of fields is then repeated per hierarchy in the analysis (i.e. density threshold). The highest column density threshold, or hierarchy, sets the final decomposition of the cloud in sub-structures. The only requirement for the aggregation of fields is their continuity in the PPV plane, therefore the shape and geometry of the identified structures is left unconstrained. Throughout the hierarchies, HiFIVe can possibly identify a transition from elongated geometries towards spherical geometries (see Sect.~\ref{subsec:MandL}). 

To describe the dense gas structures in the EMERGE Early ALMA Survey, and in particular, their radial properties (see next Section), we determined an axis for each structure and treated them as filamentary. This axis is defined as follows \citep[see][for a full discussion]{hacar13}: (a) the algorithm performs a linear fit to the structure, weighted by the column density squared, in the direction where it is most extended (i.e. with the highest variance); (b) the algorithm then samples this line into perpendicular cuts, equally spaced by $2\times\theta_\mathrm{beam}$; (c) for each of these cuts, the algorithm calculates the average coordinates of the axis knots, in arcsec, weighted again by the column density squared. The length of the identified structure is then computed as the length of this broken line.

\subsection{New implementations: the column density thresholds}\label{subsec:oldvsnew}

The hierarchical organisation of the dense gas in our targets is explored by HiFIVe through the sampling of progressively higher column densities (as proxy of the gas volume densities). Already from the previous application of the algorithm in OMC-1 and OMC-2, the integrated intensity of N$_2$H$^+$ was proved to correlate with the column density of H$_2$ \citep{hacar18}. Thanks to the calibration in Eq.~(\ref{eq:caln2hp}) we can directly identify the structures at a chosen column density regime. Once a threshold is selected, the algorithm identifies the core structures, coherent in both space and velocity (see Sect.~\ref{subsec:identification}). Those core structures composed by a number of fields larger than \texttt{N$_\mathrm{min}$} (see Sect.~\ref{sec:parexpl}) proceed to the FoF association process with the neighbouring points (see Sect.~\ref{subsec:identification}. The process is repeated per column density threshold, always keeping track of the structures identified in the previous step, i.e. the newly identified velocity-coherent structure are always enclosed within one of those determined in the previous iteration.

The description of the dense gas in OMC-1, OMC-2 and five additional regions poses new challenges, one of them being the large dynamic range in column density. As seen in Sect.~\ref{subsec:densegasprop}, the calibration in Eq.~(\ref{eq:caln2hp}) results in almost three orders of magnitude variation of N(H$_2$) in our survey (see Fig.~\ref{fig:cumulN}). The previous implementation of HiFIVe for the study of OMC-1 and OMC-2 relied on three column density thresholds, [26, 44, 66]~A$_\mathrm{v}$ \citep[$1~\mathrm{A_v} = 0.94\times10^{21}~\mathrm{cm}^{-2}$;][]{bohlin78}. As shown in Fig.~\ref{fig:cumulN} (grey lines), these thresholds cannot efficiently sample the whole column density range. In particular, they cannot describe properly its two extremes represented by NGC~2023, OMC-4 South, with $\mathrm{N(H_2)}\lesssim10^{23}~\mathrm{cm^{-2}}$ for all fields, and Flame Nebula, OMC-1, which instead show peaks above the theoretical threshold for high-mass star formation \citep{krum08}.

The first change to the previous implementation of HiFIVe is therefore a new set of column density thresholds. These are selected from 20~A$_\mathrm{v}$ to 120~A$_\mathrm{v}$ equally spaced by 20~A$_\mathrm{v}$ (i.e. [20, 40, 60, 80, 100, 120]~A$_\mathrm{v}$). The new dynamic range samples efficiently column densities within $10^{22}-10^{23}$~cm$^{-2}$, which account for $\sim80\%$ of the fields in the survey. In addition, the lowest threshold (20~A$_\mathrm{v}$) allowed us to recover $\gtrsim90\%$ of the fields in NGC~2023 within structures, while the highest threshold (120~A$_\mathrm{v}$) allowed us to explore the material at extreme densities in OMC-1/-2/-3 and Flame Nebula.

\subsection{New implementations: the integration of FilChap and radial profile fitting}\label{subsec:filchapimplement}

The second change to the previous implementation of HiFIVe is the fitting of the radial profiles. The large statistics in the EMERGE Early ALMA Survey required a systematic way of fitting these radial profiles. Towards this end, we implemented the automatic fitting routine FilChap \citep{suri19} in HiFIVe. FilChap performs a radial sampling on the selected map (the $\mathrm{N(H_2)}$ map in our analysis) along the specified axis. The single radial cuts are fitted independently or averaged into a single radial profile to then perform the fit. This last operation is carried out by the routine using four different approaches: the moments ($\mathrm{2^{nd}}$, $\mathrm{3^{rd}}$ and $\mathrm{4^{th}}$) of the radial profile, the Gaussian fit, the Plummer fits with fixed $p$-value (i.e. $p=2$ and $p=4$). Given the high-contrast, sharp profiles shown by the dense gas across the survey, we opted for the Gaussian as fitting function within FilChap to carry out the analysis. Although undoubtedly complex in both number of peaks and shape, an additional $\chi^2$ test on the sample of radial cuts suggests the Gaussian function to be the best fitting profile for our structures (see Paper IV for the full discussion).

FilChap has been previously employed as radial profile fitting routine with different filament-finding algorithms \citep[see e.g.][]{howard19}. Its implementation with HiFIVe therefore required changes to the publicly available version of the routine\footnote{\url{https://github.com/astrosuri/filchap}}.
A thorough discussion on these changes and on the parameters set in the routine to perform the fitting is given in the paper of this series which focuses on the radial profiles (Paper IV). We report here only the main adaptations: (a) FilChap takes as input both the column density and the temperature maps. The first one is used to determine the average density profile to fit along the axis of each structure; the second one is sampled along the same radial cut to give an estimate of the temperature gradient; (b) since we have already removed the floor value in Eq.~(\ref{eq:caln2hp}), no baseline subtraction has been performed during the fitting. The inclusion of FilChap gives us a systematic and direct measure of the peak column density \citep[in opposition to previous studies, e.g.][]{hacar18,suri19}, and of the width of the dense gas structures identified by HiFIVe.

\subsection{Comparison with the results in OMC-1 and OMC-2}\label{subsec:comparhacar18}

\begin{table}[tbp]
    \caption{Comparison between the results obtained with different implementations of HiFIVe and FilChap (see Sect.~\ref{subsec:oldvsnew}) in OMC-1 and OMC-2.}
    \centering
    \begin{tabular*}{\linewidth}{c| c@{\hspace{1\tabcolsep}}c c@{\hspace{0.8\tabcolsep}}c}
        \hline
        Source & \multicolumn{4}{c}{Parameter} \\ \hline
         & $\sigma_\mathrm{nt}/c_\mathrm{s}$ & $M/L$ & $FWHM$ & structures \\
        & [] & [M$_{\odot}$~pc$^{-1}$] & [pc] & [\#] \\
        \hline \\ [-2ex]
        & \multicolumn{4}{c}{Table~2, \citet{hacar18}} \\ [1.5ex]
        OMC-1 & $0.89\pm0.28$ & $24\pm22$\tablefootmark{*} & $\sim0.02-0.05$ & 28 \\ [0.7ex]
        OMC-2 & $0.65\pm0.13$ & $31\pm25$\tablefootmark{*} & $\sim0.02-0.05$ & 27 \\ [0.7ex]
        \hline \\ [-2ex]
        & \multicolumn{4}{c}{this work, [26, 44, 66]~A$_\mathrm{v}$} \\ [1.5ex]
        OMC-1 & $0.99^{+0.15}_{-0.19}$ & $25^{+14}_{-5}$ & $0.038^{+0.009}_{-0.010}$ & 27 \\ [0.7ex]
        OMC-2 & $0.60^{+0.14}_{-0.17}$ & $35^{+16}_{-14}$ & $0.046^{+0.015}_{-0.015}$ & 30 \\ [0.7ex]
        \hline \\ [-2ex]
        & \multicolumn{4}{c}{this work, [20, 40, 60, 80, 100, 120]~A$_\mathrm{v}$} \\ [1.5ex]
        OMC-1 & $0.99^{+0.12}_{-0.12}$ & $27^{+30}_{-8}$ & $0.034^{+0.010}_{-0.007}$ & 30 \\ [0.7ex]
        OMC-2 & $0.66^{+0.11}_{-0.13}$ & $40^{+37}_{-25}$ & $0.043^{+0.018}_{-0.009}$ & 46 \\ [0.7ex]
        \hline
    \end{tabular*}
    \label{tab:compartab}
    \tablefoot{
    The errors for our results have been computed as Inter-Quartile Range (IQR). \\
    \tablefoottext{*}{The line masses from \citet{hacar18} have been multiplied by a factor 2.81/2.33 to account for the different $\mu_\mathrm{H_2}$ between the two works.}
    }
\end{table}

A mandatory step in our work is to check the consistency of our new results with those derived in OMC-1 and OMC-2 by \citet{hacar18} using a previous realisation of HiFIVe. In the process, our main concern is to explore whether the new implementation produces consistent results or significant changes compared to the previous one.

As the main results of our analysis, we studied the number of identified structures, their kinematics (as $\sigma_\mathrm{nt}/c_\mathrm{s}$), their line masses (as $M/L$), their widths (as Full Width at Half Maximum, $FWHM$) using both sets of column density thresholds ([26, 44, 66]~A$_\mathrm{v}$, [20, 40, 60, 80, 100, 120]~A$_\mathrm{v}$) in our new implementation. While the parameters are described in full in the upcoming sections, we present a brief comparison in Table~\ref{tab:compartab}. Using [26, 44, 66]~A$_\mathrm{v}$, all the parameters are in excellent agreement between the two studies with the values consistent within errors. An exact matching of the results is not achieved, although expected from the use of HiFIVe directly onto the spectra and the radial profile fit with a Gaussian function. The reasons behind these mild discrepancies are the following: (a) the calibration in Eq.~(\ref{eq:caln2hp}) slightly differs from the original work, with its floor term now removed (see Sect.~\ref{subsec:densegasprop}) and the temperature map now determined using HNC/HNC instead of NH$_3$; (b) the linking metric now includes an additional $2d/\theta_\mathrm{beam}$ factor which slightly modifies the association between fields in the velocity plane; (c) the introduction of FilChap and change in fitting parameter (N(H$_2$) instead of $I(\mathrm{N_2H^+})$) alters the average widths, although only marginally. We also include the results obtained with [20, 40, 60, 80, 100, 120]~A$_\mathrm{v}$, which are also in excellent agreement with those of \citep{hacar18} and will be explored extensively in the following Sections. A full discussion on the change in results with other possible setup choices is instead given in Appendix~\ref{sec:parexpl}.

Overall, the setup choice, while arbitrary, does not lead to significant changes in the physical properties of the structures, neither in OMC-1 and OMC-2, nor across the survey (see Appendix~\ref{sec:parexpl}). Our results and conclusions prove robust thanks to the statistical significance of the sample, especially when it comes to the width of the structures identified. Thus, the EMERGE Early ALMA Survey presented in this pilot study comes as an extension of the analysis in OMC-1 and OMC-2 to the 5 additional regions in the survey.

\section{Parameter exploration}\label{sec:parexpl}

\begin{figure*}[tbp]
\centering
\includegraphics[width=0.99\linewidth]{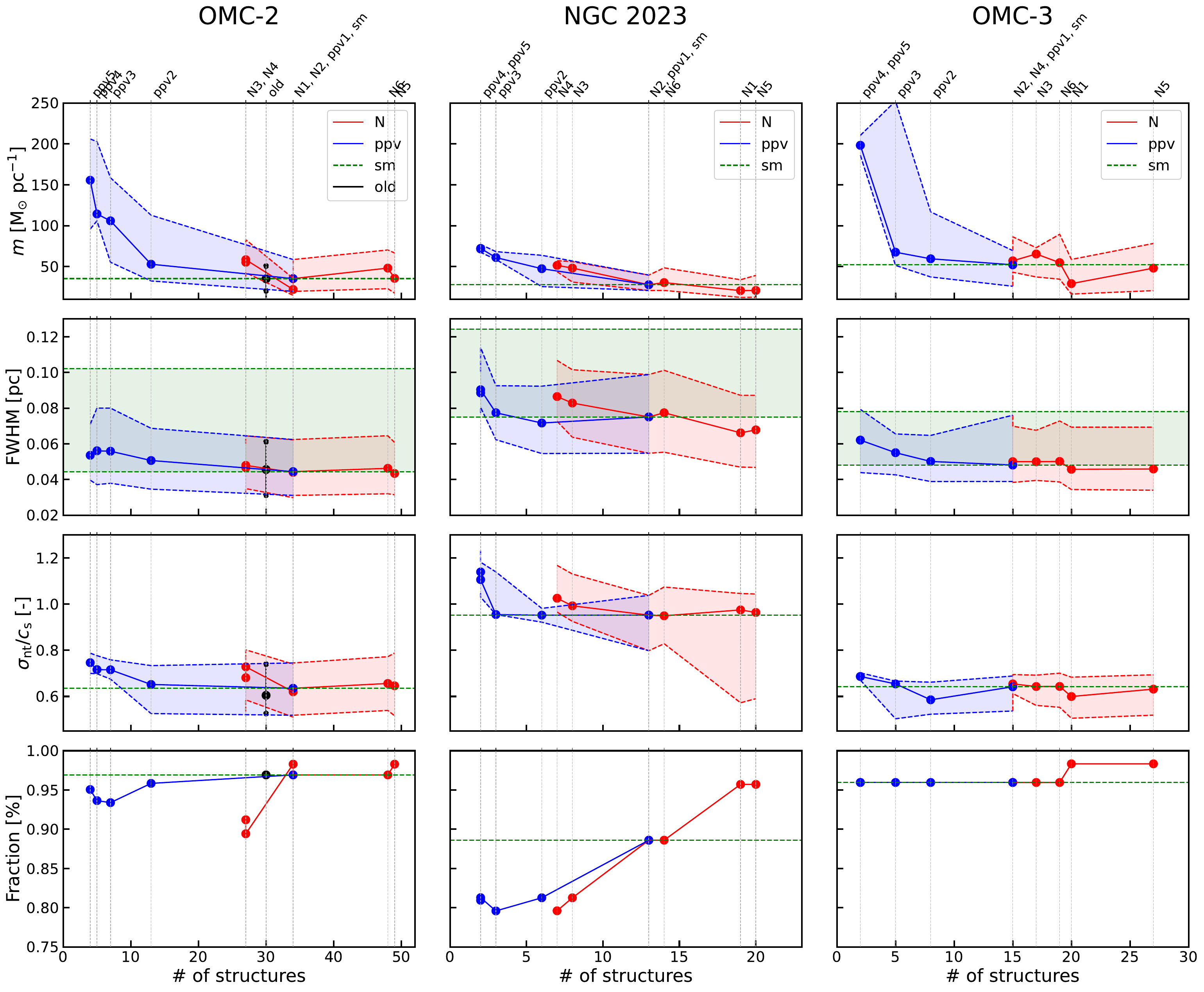}
\caption{Properties explored in OMC-2, NGC~2023 and OMC-3 when varying the setup used for the analysis (see Table~\ref{tab:setupstab} and Sect.~\ref{sec:analysis}). From top to bottom: line mass ($m$), width (FWHM), non-thermal motions ($\sigma_\mathrm{nt}/c_\mathrm{s}$), and fraction of fields recovered within fibers. The different setups are colour-coded based on the parameter varied and on the top axis there are the corresponding labels (see Table~\ref{tab:setupstab}).}
\label{fig:paramexpl}
\end{figure*}

\begin{table}[tbp]
    \caption{Different inputs given to HiFIVe and FilChap to perform the analysis (see Fig.~\ref{fig:paramexpl}).}
    \centering
    \begin{tabular*}{\linewidth}{c | c c c}
        \hline
         & \multicolumn{3}{c}{Parameter} \\ \hline
        Setup  & N(H$_2$) & mult., \texttt{Nmin} & smooth \\
        & [A$_\mathrm{V}$] & [\#, pxls] & [pxls] \\
        \hline \\ [-2ex]
        N1 & [15, 35, 55] & 1, 10 & 2 \\ [0.7ex]
        N2 & [30, 50, 70] & 1, 10 & 2 \\ [0.7ex]
        N3 & [45, 65, 85] & 1, 10 & 2 \\ [0.7ex]
        N4 & [60, 80, 100] & 1, 10 & 2 \\ [0.7ex]
        N5 & [15, 35, 55, 75, 95, 115] & 1, 10 & 2 \\ [0.7ex]
        N6 & [30, 50, 70, 90, 110, 130] & 1, 10 & 2 \\ [0.7ex]
        \hline
        \hline \\ [-2ex]
        ppv1 & [30, 50, 70] & 1, 10 & 2 \\ [0.7ex]
        ppv2 & [30, 50, 70] & 2, 26 & 2 \\ [0.7ex]
        ppv3 & [30, 50, 70] & 3, 50 & 2 \\ [0.7ex]
        ppv4 & [30, 50, 70] & 4, 82 & 2 \\ [0.7ex]
        ppv5 & [30, 50, 70] & 5, 122 & 2 \\ [0.7ex]
        \hline
        \hline \\ [-2ex]
        sm1 & [30, 50, 70] & 1, 10 & 2 \\ [0.7ex]
        sm2 & [30, 50, 70] & 1, 10 & 4 \\ [0.7ex]
        sm3 & [30, 50, 70] & 1, 10 & 6 \\ [0.7ex]
        sm4 & [30, 50, 70] & 1, 10 & 8 \\ [0.7ex]
        sm5 & [30, 50, 70] & 1, 10 & 10 \\ [0.7ex]
        \hline
        \hline \\ [-2ex]
        original & [26, 44, 66] & 1, 10 & 2 \\ [0.7ex]
        \hline
    \end{tabular*}
    \label{tab:setupstab}
\end{table}

The complexity of the dense gas morphology and dynamic range in column density cannot be fully captured by exploring a single density regime. This is the reason behind the hierarchical analysis of OMC-1 and OMC-2 done using HiFIVe \citep{hacar18}. The EMERGE Early ALMA Survey faces the challenge of studying 5 additional regions with an even broader dynamic range in column density compared to OMC-1 and OMC-2 alone. Because of the larger sample, we increase both the range and number of column density thresholds to perform the analysis (see Sect.~\ref{sec:analysis}). We have run the analysis for several different setups to assess the reliability of the results presented and in comparison with the previous analysis in OMC-1 and OMC-2 (see also Sect.~\ref{subsec:comparhacar18}).

Instead of presenting the results for the whole survey, we will focus the discussion on OMC-2, NGC~2023 and OMC-3, which we consider most representative of the different physical properties across the sample. OMC-2 has the largest dense gas mass ($\sim350$~M$_{\odot}$, see Table \ref{tab:gen_prop}) and the highest level of complexity, comprising 46 velocity-coherent fibers in the region. In addition, OMC-2 has already been studied using HiFIVe, therefore we can directly compare the results obtained with the new implementation of HiFIVe+FilChap (see also Sect.~\ref{subsec:comparhacar18}). NGC~2023 has a more diffuse and fluffy appearance of the dense gas, which, in fact, accounts for the lowest column densities across the survey (see Fig.~\ref{fig:cumulN}). As a consequence, only a small fraction of the total mass in the sample is enclosed within NGC~2023 ($\sim65$~M$_{\odot}$). OMC-3 has the second highest dense gas mass in the survey ($\sim260$~M$_{\odot}$), resulting from column densities $\gtrsim10^{23}$~cm$^{-2}$ along its main prominent crest (see Fig.~\ref{fig:NH2}). All three regions sample efficiently the column density plane, show a different dense gas morphology, even if still highly structured (see Sect.~\ref{sec:densegas}), but no extreme temperatures, such as those in OMC-1 and Flame Nebula.

When exploring the different setups for the analysis, our interest is focused on how they affect our main conclusions. Thus, we want to assess any of the following changes: first, any change in the classification of our velocity-coherent structures as dense fibers through their line mass and velocity dispersion; second, any significant change of the fiber widths in each region; finally, any change in the percentage of fields recovered by the analysis, in particular, if the setup is able to recover $\gtrsim90\%$ of the fitted spectra within structures in all the regions composing the survey.

The main changes from one setup to the other concern the column density thresholds and the linking metric for HiFIVe, and the smoothing factor to calculate the average radial profile for FilChap. In Sect.~\ref{sec:analysis}, we discussed the final column density thresholds for the analysis ($20-120$~A$_\mathrm{V}$, equally spaced by 20~A$_\mathrm{V}$). The linking metric in the Position-Position-Velocity (PPV) plane is the same used by \citet{hacar18}: (a) $2\times\theta_\mathrm{beam}$ in angular coordinates, corresponding to 10 points (\texttt{Nmin}) as minimum number of Friends to define a structure; (b) $\mathrm{FWHM}/2\times 2d/\theta_\mathrm{beam}$, in the velocity space for a Nyquist sampled spectrum. The smoothing is operated by FilChap in the sampling of the column densities along each cut perpendicular to the axis. In our final setup, we chose a \texttt{smooth} = 2, an average every two pixels, meaning each $\theta_\mathrm{beam}$ in our maps. Table \ref{tab:setupstab} shows all the different setups explored as part of this analysis. In the following we will discuss the results obtained and compare the line mass ($m$), the width ($FWHM$), the non-thermal motions ($\sigma_\mathrm{nt}/c_\mathrm{s}$) and the fraction of recovered points as key results of our analysis.

We tested the setup from \citet{hacar18} (original in Table \ref{tab:setupstab}) onto OMC-2 with our analysis. The results are shown in Fig.~\ref{fig:paramexpl} (left panels) for the different physical parameters discussed. The median line mass and non-thermal component ($35^{+16}_{-14}$~M$_{\odot}$~pc$^{-1}$, $0.60^{+0.14}_{-0.17}$) are in excellent agreement with those determined by Hacar et al. (see their Table 2), as expected since we employ HiFIVe in our analysis. The widths ($0.046^{+0.015}_{-0.015}$~pc) are also consistent with those previously determined in OMC-2. The agreement of these widths is not as good as for $m$ and $\sigma_\mathrm{nt}/c_\mathrm{s}$, however, we can ascribe these small differences to the integration of FilChap and the fitting performed directly onto the column density map (see also Sect.~\ref{subsec:comparhacar18}).

We try different sets of column densities varying the number of thresholds and the lowest value from which the selection starts (see setups N1-N6 in Table \ref{tab:setupstab}). Figure~\ref{fig:paramexpl} shows that, independently of the region, varying the column density thresholds has little effect on the physical parameters ($m$, $\sigma_\mathrm{nt}/c_\mathrm{s}$, $FWHM$), whose variation remains confined within a factor of 2. We note that a higher number of thresholds leads, in general, to a higher number of structures (see N5 and N6), as long as the lowest threshold allows to recover $\gtrsim90\%$ of the points (see the lowest panels in Fig.~\ref{fig:paramexpl}). Given the stability of the physical parameter recovered, our choice for the lowest column density threshold was dictated by the number of points recovered in NGC~2023, while their number by the complex dense gas structure seen in OMC-2.

When studying the PPV plane, we explore increasing linking metrics (through a multiplier, mult.; see setups ppv1-ppv5 in Table \ref{tab:setupstab}) and, as a consequence, increasing minimum number of Friends (\texttt{Nmin}) to determine a velocity-coherent structure. Figure~\ref{fig:paramexpl} shows the mild variations in both $FWHM$ and $\sigma_\mathrm{nt}/c_\mathrm{s}$ depending on the setup for all three regions. A higher linking metric results in a lower number of structures with increasing width and velocity dispersion overall. Still, the internal motions of the structures remain sub-sonic and the widths grow barely a factor of 2 for NGC~2023 (central panels), while remain remarkably stable for OMC-2 (left panels) and OMC-3 (right panels). $m$ is the parameter that shows the largest variability, as expected, progressively increasing with the linking metric. As the number of structures reduces, we eventually approach the total line mass for the region (see Sect.~\ref{subsec:framework} and Fig.~\ref{fig:mlinvssurf}). Again, the linking length choice is driven mostly by the ability to recover a $\gtrsim90\%$ fraction of points in velocity-coherent structures, which is achieved in NGC~2023 only for the lowest linking length.

The last parameter to explore is the smoothing factor applied by FilChap to the radial profile before its fitting (smooth; see setups sm1-sm5 in Table~\ref{tab:setupstab}). The parameter was devised to help in fitting profiles averaged across a segment or the whole filament axis and sampled along large radial distances. These profiles could in fact possess several scattered and isolated peaks, leading to a poor convergence of the fit. Figure~\ref{fig:paramexpl} shows the progressive increase of FWHM for a smoothing parameter going from 2 to 10 (see Table \ref{tab:setupstab}). The variation is particularly marked for OMC-2 and NGC~2023 (left and central panels) where the extended dense gas structures get smoothed into a single profile. OMC-3 is less affected as its sharp features do not mix even when smoothed. By considering single radial cuts along the axis, sampled within $\pm0.2$~pc from the axis knot, we already obtain relatively independent and richly sampled ($\theta_\mathrm{beam}\sim0.009$~pc) radial profiles. We therefore apply a natural smoothing of 2 pixels, a single beam of our Nyquist-sampled maps.

\end{appendix}

\end{document}